\documentclass[
  reprint,
  amsmath,amssymb,
  aps,
  prd,
  longbibliography,
  nofootinbib,
]{revtex4-2}

\usepackage{graphicx}
\usepackage{bm}
\usepackage{mathtools}
\usepackage{physics}
\usepackage{siunitx}
\usepackage{booktabs}
\usepackage{multirow}
\usepackage[dvipsnames]{xcolor}
\usepackage{orcidlink}

\newcommand{\tk}{\tilde{k}}
\newcommand{\tw}{\tilde{\omega}}
\newcommand{\twR}{\tilde{\omega}_R}
\newcommand{\tgamma}{\tilde{\gamma}}
\newcommand{\tncr}{\tilde{n}_{cr}}
\newcommand{\tvcr}{\tilde{v}_{cr}}
\newcommand{\tp}{\tilde{p}}

\newcommand{\calW}{\mathcal{W}}
\newcommand{\calH}{\mathcal{H}}
\newcommand{\run}[1]{\mathcal{R}#1}
\newcommand{\avg}[1]{\left\langle #1 \right\rangle}

\DeclareMathOperator{\sgn}{sgn}

\begin{document}

\title{Importance of the Resonant Cosmic-Ray Streaming Instability\\ Upstream of Collisionless Shocks}

\author{Bricker Ostler\,\orcidlink{0000-0003-4912-0161}}
\affiliation{Department of Physics, The University of Chicago, 5720 S Ellis Ave, Chicago, IL 60637, USA}

\author{Benedikt Schroer\,\orcidlink{0000-0002-4273-9896}}
\affiliation{Department of Astronomy and Astrophysics, The University of Chicago, 5640 S Ellis Ave, Chicago, IL 60637, USA}
\affiliation{Leibniz-Institut f\"ur Astrophysik Potsdam, An der Sternwarte 16, 14482 Potsdam, Germany}

\author{Damiano Caprioli\,\orcidlink{0000-0003-0939-8775}}
\affiliation{Department of Astronomy and Astrophysics, The University of Chicago, 5640 S Ellis Ave, Chicago, IL 60637, USA}
\affiliation{Enrico Fermi Institute, The University of Chicago, 933 E 56th St, Chicago, IL 60637, USA}

\date{\today}

\begin{abstract}
Cosmic rays (CRs) escaping collisionless shocks form a dilute, relativistic population whose current amplifies the upstream magnetic field, a process widely attributed to the non-resonant (Bell) instability. Solving the dispersion relation for both cold and finite-spread CR distributions, we show that for the maximum-energy escaping population, the resonant mode can outgrow the non-resonant mode. At slower shocks, this dominance persists for broader CR distributions, and sufficient pitch-angle broadening can even stabilize the non-resonant mode while leaving the resonant mode unstable. Relativistic hybrid particle-in-cell simulations confirm the cold-beam linear theory predictions and, in the nonlinear regime, saturate at $\delta B/B_0 \sim 1$ with significant pitch-angle redistribution. Neglecting the resonant instability thus underestimates magnetic field growth at the very scale needed to confine the highest-energy escaping CRs.
\end{abstract}

\maketitle

\section{Introduction}

Collisionless shocks, especially those associated with supernova remnants (SNRs), are believed to be capable of accelerating cosmic rays (CRs) up to energies near the ``knee'' of the CR energy spectrum around $1$ PeV via diffusive shock acceleration (DSA) \cite{Blasi2019, Bell1978, Blandford+1978}. In the standard picture, a small fraction of initially thermal ions swept up by the shock are injected into the acceleration process and undergo repeated scatterings across the shock between the upstream and downstream flows. Eventually, they form a non-thermal population whose quasi-isotropic distribution function at the shock front approximately follows a power law in momentum.

Sufficiently far upstream, the magnetic fluctuations that confine particles via pitch-angle scattering become too weak to confine the most energetic CRs. These CRs therefore propagate onward as a dilute, relativistic, forward-beamed population. Their streaming amplifies the background magnetic field via the resonant or non-resonant (Bell) instability \cite{Bell1978, Bell2004, AmatoBlasi2009}, confining future particles of smaller or comparable energy. By enabling further energy gain via DSA, this confinement leads to a maximum-achievable energy that increases with time.

The non-resonant instability, driven by the return current carried by the background electrons, has a growth rate that is relatively insensitive to the CR distribution provided the CRs are unmagnetized \cite{Zacharegkas+2024, Das+2026}. In contrast, the resonant instability, driven by gyroresonant coupling between the CRs and magnetic fluctuations, changes significantly as the CR distribution broadens. For the fastest-growing mode, CRs with different momenta and pitch angles differ in how closely they satisfy the gyroresonance condition. A sufficiently narrow, beam-like distribution gives rise to a \textit{reactive} resonant instability when the particle-to-particle spread in resonance mismatch is smaller than the mode's growth rate, allowing the beam to respond coherently \cite{Melrose1986, Lyutikov1999}. When this spread exceeds the growth rate, the instability becomes \textit{kinetic}, with mode growth driven primarily by the CRs closest to exact gyroresonance \cite{KulsrudPearce69, Zweibel1979, Achterberg1983, Holcomb+2019}. Even in this kinetic limit, the CR drive can substantially modify the mode beyond a simple Alfv\'enic perturbation.

In this article, we investigate the competition between the resonant and non-resonant modes in the far upstream of a collisionless shock. We find that, for parameters typical of SNRs, a sufficiently beam-like escaping CR population induces a resonant instability that dominates mode growth. This instability is reactive in the cold-beam limit and, particularly at slower shocks, remains dominant once finite momentum and pitch-angle spreads make the instability kinetic. Since the fastest-growing mode is gyroresonant with the escaping CRs, it amplifies magnetic fluctuations capable of scattering the very particles driving it.

The paper is organized as follows. Section~\ref{sec:general-dispersion-relation} derives the cold-beam dispersion relation for transverse modes propagating parallel to the background magnetic field and shows that, at fiducial SNR parameters, the resonant mode dominates magnetic field growth. Section~\ref{sec:sensitivity-res-growth} quantifies the reactive-to-kinetic transition as momentum and pitch-angle spreads increase, Sec.~\ref{sec:hybrid-pic-simulations} tests the cold-beam predictions against relativistic hybrid particle-in-cell simulations, and Sec.~\ref{sec:discussion} places the results in the context of previous work.

\section{General Dispersion Relation} \label{sec:general-dispersion-relation}

Consider a plasma consisting of cold beams of non-relativistic electrons and ions (protons), and potentially relativistic CRs (protons), where $n_s$ and $v_s$ denote the number density and drift velocity of species $s$. This composition describes the far-upstream region of the shock where the confined, quasi-isotropic population is rare, leaving only the escaping CR population. We take the plasma to be in a stationary, homogeneous, quasi-neutral equilibrium with a uniform background magnetic field $\vb{B}_0=B_0\vu{x}$. Consistency with the Amp\`ere--Maxwell equation then requires the total equilibrium current to vanish. Adopting the ion bulk-rest frame for the dispersion-relation analysis, quasi-neutrality and zero net current yield

\begin{equation}\label{eq:quasi-neutrality}
    n_{e} = n_i + n_{cr} \qc
    v_{e} = \frac{n_{cr}}{n_e} v_{cr}.
\end{equation}

We focus on the properties of transverse modes with wavevectors parallel to $\vb{B}_0$, namely $\vb{k} = k_x\vu{x}$ and $k\equiv k_x>0$.

In the circular basis, $\vu{e}_\pm = (\vu{y} \pm i\vu{z})/\sqrt{2}$, the two transverse polarizations decouple, resulting in two dispersion relations,

\begin{equation}\label{eq:general-dispersion-relation}
    \frac{k^2 c^2}{\omega^2} = 1 + \sum_s \chi_s^{\sigma},
\end{equation}

\noindent where $\chi_s^{\sigma}$ is the susceptibility of species $s$ for circular basis sign $\sigma = \pm 1$ and $\omega = \omega_R + i\gamma$ \cite{Stix1992, Gary1993}. We consider modes with $\abs{\omega} \ll k c$, allowing the factor of ``1'' arising from the displacement current to be neglected. Using the ion inertial length $d_i = c/\omega_{pi}$, Eq.~\eqref{eq:general-dispersion-relation} may be rewritten as

\begin{equation} \label{eq:specific-dispersion-relation}
    k^2 d_i^2 = \frac{\omega^2}{\omega_{pi}^2} \pqty{\chi_e^\sigma + \chi_i^\sigma + \chi_{cr}^\sigma}.
\end{equation}

Each susceptibility is given by

\begin{equation} \label{eq:susceptibility}
    \chi_s^{\sigma} = -\frac{\omega_{ps}^2}{\Gamma_{s}\omega^2} \frac{\omega - kv_{s}}{\omega - kv_{s} + \sigma \Omega_s/\Gamma_{s}},
\end{equation}

\noindent where $\omega_{ps} = \sqrt{4\pi q_s^2 n_s / m_s}$ is the plasma frequency, $n_s$ the number density in the ion bulk-rest frame, $\Omega_s = q_s B_0 / (m_s c)$ the cyclotron frequency, and $\Gamma_{s}$ the Lorentz factor at drift velocity $v_s$.

We restrict our attention to low-frequency growing modes satisfying $\abs{\omega} \ll \Omega_i$ and $\abs{\omega'} \ll \abs{\Omega_e}$, where $\omega' = \omega - kv_e$ is the mode frequency as seen in the electron bulk-rest frame. Using $1/(1+x) = 1 - x + \order{x^2}$, the ion susceptibility retained to 1\textsuperscript{st} order in $\omega/(\sigma\Omega_i)$ is

\begin{equation}
    \chi_i^\sigma = -\frac{\omega_{pi}^2}{\omega^2}\frac{\omega}{\omega + \sigma\Omega_i} \approx -\frac{\omega_{pi}^2}{\omega^2}\pqty{\frac{\omega}{\sigma \Omega_i} - \frac{\omega^2}{\Omega_i^2}}.
\end{equation}

Similarly, the electron susceptibility retained to 0\textsuperscript{th} order in $\omega'/(\sigma\Omega_e)$ is

\begin{equation}
    \chi_e^\sigma = -\frac{\omega_{pe}^2}{\omega^2} \frac{\omega'}{\omega' + \sigma \Omega_e} \approx -\frac{\omega_{pe}^2}{\omega^2} \frac{\omega - kv_e}{\sigma \Omega_e}.
\end{equation}

Using Eq.~\eqref{eq:quasi-neutrality} and $\omega_{ps}^2 / \omega_{pi}^2 = q_s n_s \Omega_s / (e n_i \Omega_i)$, Eq.~\eqref{eq:specific-dispersion-relation} then simplifies to

\begin{align} \label{eq:dispersion-relation-simplified}
k^2 d_i^2
&=
\frac{\omega^2}{\Omega_i^2}
+
\frac{n_{cr}}{n_i}
\frac{\omega-kv_{cr}}{\sigma\Omega_i}
\nonumber\\
&\quad
-
\frac{1}{\Gamma_{cr}}
\frac{n_{cr}}{n_i}
\frac{\omega-kv_{cr}}
{\omega-kv_{cr}+\sigma\Omega_{cr}/\Gamma_{cr}} ,
\end{align}

\noindent which may be written in the factored form

\begin{equation} \label{eq:dispersion-relation-factored}
    k^2 v_{A0}^2 - \omega^2 = \sigma \Omega_i\frac{n_{cr}}{n_i}\frac{\pqty{\omega - k v_{cr}}^2}{\omega - k v_{cr} + \sigma \Omega_{cr} / \Gamma_{cr}}
\end{equation}

\noindent for $\Omega_{cr} = \Omega_i$ and $v_{A0} = B_0 / \sqrt{4\pi m_i n_i} = d_i\Omega_i$. In what follows, we detail simple approximate solutions to this dispersion relation applicable to SNR environments for either sign of $\sigma$, or equivalently either helicity $\calH = \sigma \sgn (k_x) = \sigma$ under our $k_x>0$ convention (Appendix~\ref{app:basis-sign-pol-hel}). The exact solutions are given in Appendix~\ref{app:exact-dr-solution}.

\subsection{Non-resonant dispersion relation} \label{sec:nonres-dr}

We now consider the right-hand helicity ($\calH = -1$) branch of Eq.~\eqref{eq:dispersion-relation-simplified}. Thus far, we have not approximated the CR susceptibility. However, for modes with wavenumbers much larger than the resonant wavenumber, the CRs will appear effectively unmagnetized; namely, 

\begin{equation} \label{eq:unmagnetized-approx}
    \abs{\omega''} = \Gamma_{cr} \abs{\omega - kv_{cr}} \gg \Omega_{cr},
\end{equation}

\noindent where $\omega''$ is the mode frequency as seen in the CR rest frame. For $\abs{\omega} \ll k v_{cr}$, this is equivalent to the spatial criterion $kr_{cr} \gg 1$ for gyroradius $r_{cr} = \Gamma_{cr} v_{cr} / \Omega_{cr}$. To 1\textsuperscript{st} order in $\Omega_{cr}/\omega''$, Eq.~\eqref{eq:dispersion-relation-factored} thus reduces to

\begin{equation} \label{eq:non-res-dispersion-relation}
    k^2 v_{A0}^2 - \omega^2 \simeq - \Omega_i\frac{n_{cr}}{n_i}\pqty{\omega - k v_{cr}} - \Omega_i^2 \frac{n_{cr}}{n_i}\frac{1}{\Gamma_{cr}}.
\end{equation}

Substituting $\omega = \omega_R + i\gamma$ into this dispersion relation and taking real and imaginary parts gives the wavenumber-independent real frequency 

\begin{equation}
    \omega_R = \frac{1}{2}\frac{n_{cr}}{n_i} \Omega_i
\end{equation}

\noindent for $\gamma \neq 0$ and growth rate

\begin{equation} \label{eq:non-resonant-gammak}
    \gamma(k) = \sqrt{\Omega_i \frac{n_{cr}}{n_i}kv_{cr} - k^2 v_{A0}^2 - \Omega_i^2\frac{n_{cr}}{n_i}\frac{1}{\Gamma_{cr}} - \Omega_i^2 \frac{n_{cr}^2}{4n_i^2}}.
\end{equation}

 The corresponding fastest-growing wavenumber $k_{\max}^{\rm nr}$ and growth rate $\gamma_{\max}^{\rm nr} \equiv \gamma\pqty{k_{\max}^{\rm nr}}$ are

\begin{equation} \label{eq:non-resonant-kfgm}
    k_{\max}^{\rm nr} d_i = \frac{\omega_R}{\Omega_i} \frac{v_{cr}}{v_{A0}} = \frac{1}{2}\frac{n_{cr}}{n_i}\frac{v_{cr}}{v_{A0}}
\end{equation}

\noindent and

\begin{equation} \label{eq:gammafgm-non-res-full}
    \gamma_{\max}^{\rm nr} = \omega_R\sqrt{\frac{v_{cr}^2}{v_{A0}^2}-1 - \frac{4n_{i}}{n_{cr}}\frac{1}{\Gamma_{cr}}}.
\end{equation}

In the limit of a sufficiently super-Alfv\'enic ($v_{cr} \gg v_{A0}$) CR species satisfying $n_{cr} \Gamma_{cr} / n_i \gg (v_{A0}/v_{cr})^2$, Eq.~\eqref{eq:gammafgm-non-res-full} reduces to 

\begin{equation} \label{eq:non-resonant-gammafgm-approx}
    \gamma_{\max}^{\rm nr} \simeq \omega_R \frac{v_{cr}}{v_{A0}} = k_{\max}^{\rm nr} v_{A0},
\end{equation}

\noindent as expected from the non-resonant (Bell) instability \cite{WinskeGary1986, Bell2004}. Since the group velocity under this approximation is $v_{g} = \partial \omega_R / \partial k = 0$, the instability is approximately stationary and purely growing in the ion bulk-rest frame. At sufficiently low CR current, the mode becomes stable across all wavenumbers (see Appendix~\ref{app:stability-non-res}). 

\subsection{Resonant dispersion relation} \label{sec:resonant-dr}

The resonant mode arises from the left-hand helicity ($\calH = +1$) branch of Eq.~\eqref{eq:dispersion-relation-simplified}. Defining the detuning $\Delta(p,\mu;k)$ of a CR with momentum $p$ and pitch-angle cosine $\mu$ from a mode with wavenumber $k$ and real frequency $\omega_R(k)$,

\begin{equation}
    \Delta(p,\mu; k) \equiv \omega_R(k) - kv(p)\mu + \frac{\Omega_{cr}}{\Gamma(p)},
\end{equation}

\noindent we then make the change of variables 

\begin{equation} \label{eq:ansatz}
    \omega = kv_{cr} - \frac{\Omega_{cr}}{\Gamma_{cr}} + \Delta_0 + i\gamma,
\end{equation}

\noindent where $\Delta_0(k) \equiv \Delta(p_{cr},1;k)$ is the detuning away from exact cold-beam resonance and $p_{cr} \equiv \Gamma_{cr} m_{cr} v_{cr}$. Substituting Eq.~\eqref{eq:ansatz} into Eq.~\eqref{eq:dispersion-relation-factored} then gives

\begin{equation} \label{eq:res-dr-approx}
    \pqty{k^2 v_{A0}^2 - \omega^2}\pqty{\Delta_0 + i\gamma} = \Omega_i \frac{n_{cr}}{n_i}\pqty{\Delta_0 + i \gamma - \frac{\Omega_{cr}}{\Gamma_{cr}}}^2.
\end{equation}

Then, we make the following assumptions about our modes which we verify \textit{a posteriori}:

\begin{equation} \label{eq:fgm-assumptions}
    \abs{\omega} \gg k v_{A0}  \qc \abs{\Delta_0 + i\gamma} \ll \frac{\Omega_{cr}}{\Gamma_{cr}}.
\end{equation}

Defining $\calW  \equiv kv_{cr} - \Omega_{cr}/\Gamma_{cr}$, these assumptions reduce Eq.~\eqref{eq:res-dr-approx} to

\begin{equation}
    \pqty{\calW + \Delta_0 + i\gamma}^2\pqty{\Delta_0 + i\gamma} = -\Omega_i^3\frac{n_{cr}}{n_i\Gamma_{cr}^2}.
\end{equation}

The imaginary part yields an expression for the growth rate,

\begin{equation}\label{eq:impart-res-approx}
    \gamma^2 = \pqty{\calW + \Delta_0}\pqty{\calW + 3\Delta_0},
\end{equation}

\noindent which, upon substitution into the real part, yields

\begin{equation}\label{eq:repart-res-approx}
    \pqty{\calW+\Delta_0}\pqty{\calW + 2\Delta_0}^2 = \Omega_i^3\frac{n_{cr}}{2n_i\Gamma_{cr}^2}.
\end{equation}

We obtain the fastest-growing wavenumber by differentiating Eq.~\eqref{eq:repart-res-approx} with respect to $k$ and using $d\calW / dk = v_{cr}$, giving

\begin{equation}
    \dv{\Delta_0}{k} = -v_{cr} \frac{3 \calW + 4\Delta_0}{5\calW + 6\Delta_0}.
\end{equation}

Plugging this into the differentiated $\gamma^2$ then yields

\begin{equation}
    \dv{\gamma^2}{k} = -\frac{2v_{cr}\calW\pqty{\calW+\Delta_0}}{5\calW + 6\Delta_0}.
\end{equation}

Since the right side of Eq.~\eqref{eq:repart-res-approx} is positive, we have $\calW+\Delta_0 > 0$. Therefore, $d\gamma^2/dk = 0$ implies $\calW = 0$, giving

\begin{equation}\label{eq:resonant-kfgm-approx}
    k_{\max}^{\rm res} = \frac{\Omega_{cr}}{\Gamma_{cr}v_{cr}} = \frac{1}{r_{cr}}. 
\end{equation}

We then immediately obtain the fastest-growing mode's growth rate and frequency detuning from Eqs.~\eqref{eq:impart-res-approx} and \eqref{eq:repart-res-approx}:

\begin{equation}\label{eq:resonant-gammafgm-delta-approx}
    \gamma_{\max}^{\rm res} = \sqrt{3}\Delta_0^{\rm max} = \frac{\sqrt{3}}{2}\pqty{\frac{n_{cr}}{n_i \Gamma_{cr}^2}}^{1/3} \Omega_i.
\end{equation}

This cube-root scaling in $n_{cr}/n_i$ is characteristic of the reactive resonant instability \cite{Gary1993, Weidl+2019, Shalaby+2021}. In contrast, the scaling becomes square-root in the strongly driven kinetic limit \cite{AmatoBlasi2009} and linear in the weakly driven Alfv\'enic limit \cite{KulsrudPearce69, Holcomb+2019}.

Notably, Eqs.~\eqref{eq:ansatz} and \eqref{eq:resonant-kfgm-approx} imply the fastest-growing mode has phase and group velocities of

\begin{equation}
    v_{\phi}^{\max} = \frac{v_{cr}}{2}\pqty{\frac{n_{cr} \Gamma_{cr}}{n_i}}^{1/3}
\end{equation}

\noindent and

\begin{equation} \label{eq:fgm-group-velocity}
    v_g^{\max} = v_{cr} + \dv{\Delta_0}{k}\eval_{\mathcal{W}=0} = \frac{v_{cr}}{3}.
\end{equation}

If we consider the fiducial parameters of a young SNR with  $v_{\rm sh} = 5000\,{\rm km\,s^{-1}}$ (see Table~\ref{tab:fiducial-snr-parameters}), we obtain $v_\phi^{\max} \simeq 93 \, v_{A0}$ and $v_g^{\max} \simeq 2\times 10^4\, v_{A0}$. This bears little resemblance to an Alfv\'enic perturbation with both velocities on the order of $v_{A0}$.

\subsection{Estimates for SNRs} \label{sec:SNR-estimate}

These approximate formulae for the fastest-growing mode are valid as long as that mode satisfies Eq.~\eqref{eq:fgm-assumptions}. Using Eqs.~\eqref{eq:resonant-kfgm-approx} and \eqref{eq:resonant-gammafgm-delta-approx}, this yields a constraint on $n_{cr} \Gamma_{cr} / n_i$:

\begin{equation} \label{eq:ordering}
    \pqty{\frac{v_{A0}}{v_{cr}}}^3 \ll \frac{n_{cr}\Gamma_{cr}}{n_i} \ll 1.
\end{equation}

We can determine whether Eq.~\eqref{eq:ordering} holds for typical SNR environments as follows. Assume that, near the shock front, the CRs follow the test-particle DSA spectrum

\begin{equation}
    f_0(\vb{x}=0,p) \equiv f_0(p) = f_0(p_{cr}) \pqty{\frac{p}{p_{cr}}}^{-q}
\end{equation}

\noindent of index $q = 3r/(r-1)$ for a shock with compression ratio $r$ and CRs with momenta $p\in\bqty{p_{\rm min}, p_{\max}}$ where $p_{\min} = m_{cr} c$ and $p_{\max} = p_{cr} \simeq \Gamma_{cr} m_{cr} c$. From Appendix~\ref{app:FEB-shock-relationship}, the escaping CR current far upstream equals its value at the shock up to corrections of $\order{v_{\rm sh}/c}$,

\begin{equation} \label{eq:Jcr-vshock-relation}
    J_{cr} = e n_{cr} v_{cr} \simeq \frac{4\pi}{q}\, e\, v_{\rm sh}\, p_{cr}^3 f_0(p_{cr}),
\end{equation}

\noindent for $q>3$ and $v_{cr}\simeq c$. Moreover, the CR scalar pressure
at the shock is

\begin{align}
    P_{cr} &= \frac{4\pi}{3}\int_{m_{cr} c}^{p_{cr}} p v f_0(p)p^2 \dd{p} \nonumber \\
           &\simeq \frac{4\pi c}{3} p_{cr}^4 f_0(p_{cr}) \Lambda,
\end{align}

\noindent where $\Lambda \equiv \int_{1/\Gamma_{cr}}^1 \tp^{3-q} \dd{\tp}$ and $\tp\equiv p/p_{cr}$. Eliminating \(f_0(p_{cr})\) and using the CR acceleration efficiency $\xi_{cr} = P_{cr} / (\rho v_{\rm sh}^2)$ for $\rho \simeq m_i n_i$ then yields

\begin{equation}
    \label{eq:shock-relation}
    \frac{n_{cr}\Gamma_{cr}}{n_i}
    =
    \frac{3}{q\Lambda}
    \pqty{\frac{v_{\rm sh}}{c}}^3
    \xi_{cr}.
\end{equation}

With the fiducial SNR parameters summarized in
Table~\ref{tab:fiducial-snr-parameters} and $v_{\rm sh}=5000\,{\rm km\,s^{-1}}$, we obtain

\begin{equation}
    \pqty\bigg{\frac{v_{A0}}{c}}^3 \sim 10^{-15} \qc \frac{n_{cr}\Gamma_{cr}}{n_i} \sim 10^{-8},
\end{equation}

\noindent well within the region of validity of Eq.~\eqref{eq:ordering}. Assuming a fixed shock speed and acceleration efficiency, spectra with indices $q>4$ \cite{Caprioli2012, Caprioli+2020} result in

\begin{equation}
    \frac{\pqty{n_{cr}/n_i}_{q>4}}{\pqty{n_{cr}/n_i}_{q=4}} = \frac{4(q-4)}{q} \frac{\ln\Gamma_{cr}}{\Gamma_{cr}^{q-4}-1},
\end{equation}

\noindent using Eq.~\eqref{eq:shock-relation}, which for $\Gamma_{cr}=10^5$ at $q=4.3$ lowers $n_{cr}/n_i$ by an order of magnitude relative to the $q=4$ case. This leaves Eq.~\eqref{eq:ordering} satisfied and, as shown below (Eq.~\eqref{eq:cold-beam-crossover-vshock}), strengthens the resonant mode's dominance over the non-resonant mode.

\begin{table}[t]
\centering
\begin{tabular}{cc}
\toprule
Parameter & Fiducial value \\
\midrule
$v_{A0}$ & $5\,{\rm km\,s^{-1}}$ \\
$\xi_{cr}$ & $0.1$ \\
$\Gamma_{cr}$ & $10^5$ \\
$q$ & $4$ \\
$n_{cr}/n_i$ & $3 \times 10^{-13}\,\pqty{v_{\rm sh} / 5000\,{\rm km\,s^{-1}}}^{3}$ \\
\bottomrule
\end{tabular}
\caption{Fiducial SNR parameters adopted in this work. The first four parameters are fixed throughout Secs.~\ref{sec:general-dispersion-relation}--\ref{sec:sensitivity-res-growth}; the value of $n_{cr}/n_i$ is then determined by Eq.~\eqref{eq:shock-relation} once $v_{\rm sh}$ is specified. The Alfv\'en speed $v_{A0}$ corresponds to an ion number density of $n_i = 1\,{\rm cm^{-3}}$ in a background magnetic field of $B_0 \simeq 2.3 \,{\rm \mu G}$.}
\label{tab:fiducial-snr-parameters}
\end{table}

\subsection{Comparison of the resonant and non-resonant modes}

\begin{figure}
\centering
\includegraphics[width=\columnwidth]{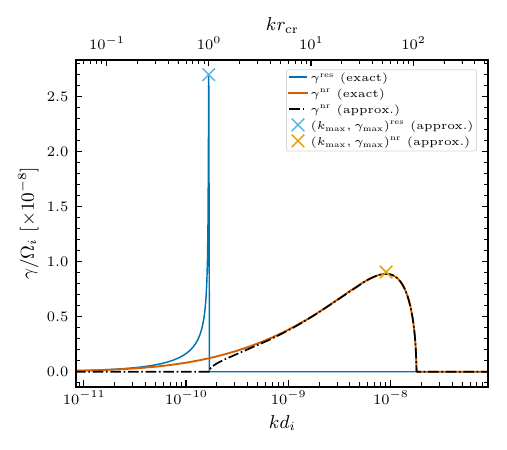}
\caption{\label{fig:growth-rate-vs-k}Growth rates vs. wavenumber for the resonant and non-resonant modes at the fiducial parameters of Table~\ref{tab:fiducial-snr-parameters} and $v_{\rm sh}=5000\,{\rm km\,s^{-1}}$. The approximate fastest-growing wavenumbers and growth rates agree well with the exact solutions, with the resonant mode dominating.}
\end{figure}

Figure~\ref{fig:growth-rate-vs-k} compares the resonant and non-resonant growth rates at the fiducial SNR parameters of Table~\ref{tab:fiducial-snr-parameters}. We plot the exact solution of Eq.~\eqref{eq:dispersion-relation-factored} for both modes alongside the unmagnetized-limit approximation for the non-resonant mode, Eq.~\eqref{eq:non-resonant-gammak}. The approximate fastest-growing wavenumbers (Eqs.~\eqref{eq:non-resonant-kfgm} and~\eqref{eq:resonant-kfgm-approx}) and growth rates (Eqs.~\eqref{eq:non-resonant-gammafgm-approx} and~\eqref{eq:resonant-gammafgm-delta-approx}) agree well with the exact solutions. As expected, the non-resonant approximation underestimates the exact growth rate at low~$k r_{cr}$, a consequence of treating the CRs as unmagnetized.

Equating the approximate maximum growth rates yields the cold-beam crossover
\begin{equation}\label{eq:cold-beam-crossover}
    \pqty{\frac{n_{cr}\Gamma_{cr}}{n_i}}_{\rm cross}
    =
    3^{3/4}\pqty{\frac{v_{A0}}{v_{cr}}}^{3/2}.
\end{equation}

\noindent The non-resonant mode dominates above this value, whereas the reactive resonant mode dominates below it. Combining this crossover condition with Eq.~\eqref{eq:shock-relation} gives the corresponding crossover shock speed,
\begin{align}
    v_{\rm sh}^{\rm cross}
    &\simeq
    \sqrt{c v_{A0}}
    \pqty{\frac{q\Lambda}{3^{1/4}\xi_{cr}}}^{1/3}
    \nonumber\\
    &\simeq
    8600\,{\rm km\,s^{-1}}
    \pqty\bigg{\frac{B_0}{2.3\,{\rm \mu G}}}^{1/2}
    \pqty\bigg{\frac{n_i}{1\,{\rm cm^{-3}}}}^{-1/4}
    \nonumber\\
    &\quad\times
    \pqty\bigg{\frac{q}{4}}^{1/3}
    \pqty\bigg{\frac{\Lambda}{\ln(10^5)}}^{1/3}
    \pqty\bigg{\frac{\xi_{cr}}{0.1}}^{-1/3}.
    \label{eq:cold-beam-crossover-vshock}
\end{align}

If CRs escaped as an idealized cold beam, the resonant mode would therefore be the dominant mode for all historical SNRs.

\section{Sensitivity of the Resonant Growth Rate to Momentum and Pitch-Angle Spread} \label{sec:sensitivity-res-growth}

In the preceding sections, we investigated the growth rate of the resonant mode for an idealized cold beam. Since realistic escaping CR distributions contain a non-zero spread in momentum and pitch angle, we quantify how this spread affects the peak growth rate. We first derive an analytic estimate for the onset of decreased growth rate for the fastest-growing mode, and then numerically solve for the fastest-growing modes using the full dispersion relation with a reasonable choice of CR distribution function. 

\subsection{Analytic estimate}

Let $\delta p = p - p_{cr}$ and $\delta \mu = \mu - 1$ denote deviations from the cold-beam momentum and pitch-angle cosine. At fixed $k$, the detuning deviation is

\begin{align}
    \delta \Delta (p,\mu;k) &\equiv \Delta(p,\mu;k) - \Delta_0(k) \nonumber \\
    &= k\pqty{v_{cr} - v\mu} + \Omega_{cr} \pqty{\frac{1}{\Gamma} - \frac{1}{\Gamma_{cr}}}.
\end{align}

Expanding about the cold-beam point gives, to first order,

\begin{align}
    \delta \Delta &\simeq \pdv{\Delta}{p}\eval_0 \delta p + \pdv{\Delta}{\mu}\eval_0 \delta \mu \nonumber \\
    &= -\pqty{\frac{kv_{cr}}{\Gamma_{cr}^2} + \frac{\Omega_{cr}}{\Gamma_{cr}}\frac{v_{cr}^2}{c^2}}\frac{\delta p}{p_{cr}} - kv_{cr} \delta \mu,
\end{align}

\noindent where $\eval_0$ denotes evaluation at $(p,\mu) = (p_{cr},1)$. Defining the detuning deviation from the cold-beam fastest-growing mode as $\delta \Delta^{\rm max} \equiv \delta \Delta (p,\mu; k_{\max}^{\rm res})$, this reduces to 

\begin{equation} \label{eq:detuning-deviation}
    \delta \Delta^{\rm max} \simeq -\frac{\Omega_{cr}}{\Gamma_{cr}} \pqty{\frac{\delta p}{p_{cr}} + \delta \mu}.
\end{equation}

Squaring Eq.~\eqref{eq:detuning-deviation} and taking the moment
$\avg{\cdot} = n_{cr}^{-1}\int_{\mathbb{R}^3} \dd[3]{p} f_{cr}\,\pqty{\cdot}$
of the resulting equation then gives the mean-squared detuning deviation,
\begin{equation}
    \avg{\pqty{\delta \Delta^{\rm max}}^2} \simeq
    \frac{\Omega_{cr}^2}{\Gamma_{cr}^2 p_{cr}^2}\avg{\delta p ^2}
    +
    \frac{\Omega_{cr}^2}{\Gamma_{cr}^2}\avg{\delta \mu^2}
    +
    \frac{2\Omega_{cr}^2}{\Gamma_{cr}^2 p_{cr}}\avg{\delta p \, \delta \mu}.
\end{equation}

Thus,
\begin{equation}
    \sigma_{\Delta^{\rm max}} \simeq
    \frac{\Omega_{cr}}{\Gamma_{cr}}
    \sqrt{\frac{\sigma_p^2}{p_{cr}^2} + \sigma_\mu^2 + \frac{2\avg{\delta p \, \delta \mu}}{p_{cr}}},
\end{equation}

\noindent where $\sigma_X \equiv \sqrt{\avg{\delta X^2}}$ is the root-mean-square (rms) deviation of $X\in\qty{\Delta^{\rm max},p,\mu}$ from its cold-beam point. 

As an order-of-magnitude criterion, a resonant mode remains reactive as long as $\sigma_{\Delta} \lesssim \gamma^{\rm res}$ \cite{Lyutikov1999, ONiel+68}. Using Eq.~\eqref{eq:resonant-gammafgm-delta-approx} for the fastest-growing reactive mode, this becomes

\begin{equation} \label{eq:dropoff-estimate}
    \sqrt{\frac{\sigma_p^2}{p_{cr}^2} + \sigma_\mu^2 + \frac{2\avg{\delta p \, \delta \mu}}{p_{cr}}} \lesssim \frac{\sqrt{3}}{2}\pqty{\frac{n_{cr}\Gamma_{cr}}{n_i}}^{1/3}.
\end{equation}

\begin{figure}
\centering
\includegraphics[width=\columnwidth]{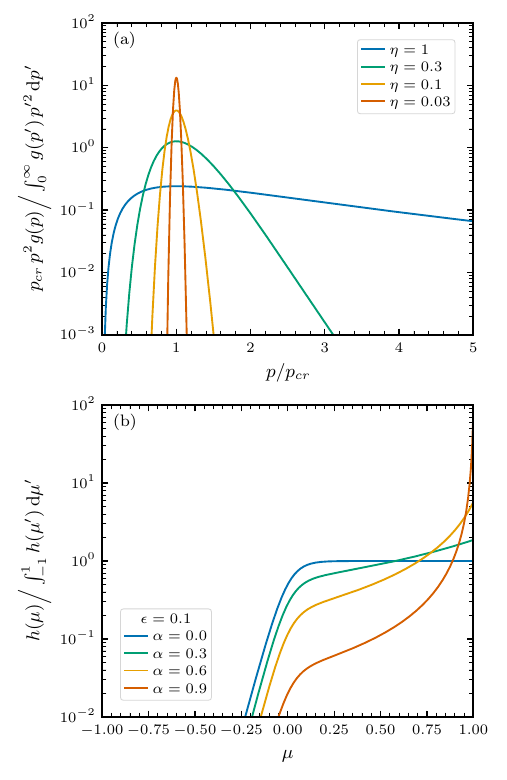}
\caption{\label{fig:cr-distribution}(a) Normalized CR momentum spectrum $(p_{cr}/\mathcal{N}_{cr}) \, dN_{cr}/dp$ where $\mathcal{N}_{cr} = \int f_{cr}(\vb{x},\vb{p})\dd[3]{x}\dd[3]{p}$. (b) Normalized CR pitch-angle cosine spectrum $(1/\mathcal{N}_{cr})\, dN_{cr}/d\mu$. The sharp cutoff at $\mu=0$ is smoothed via Eq.~\eqref{eq:heaviside-smoothing}.}
\end{figure}

\begin{figure*}
\centering
\includegraphics{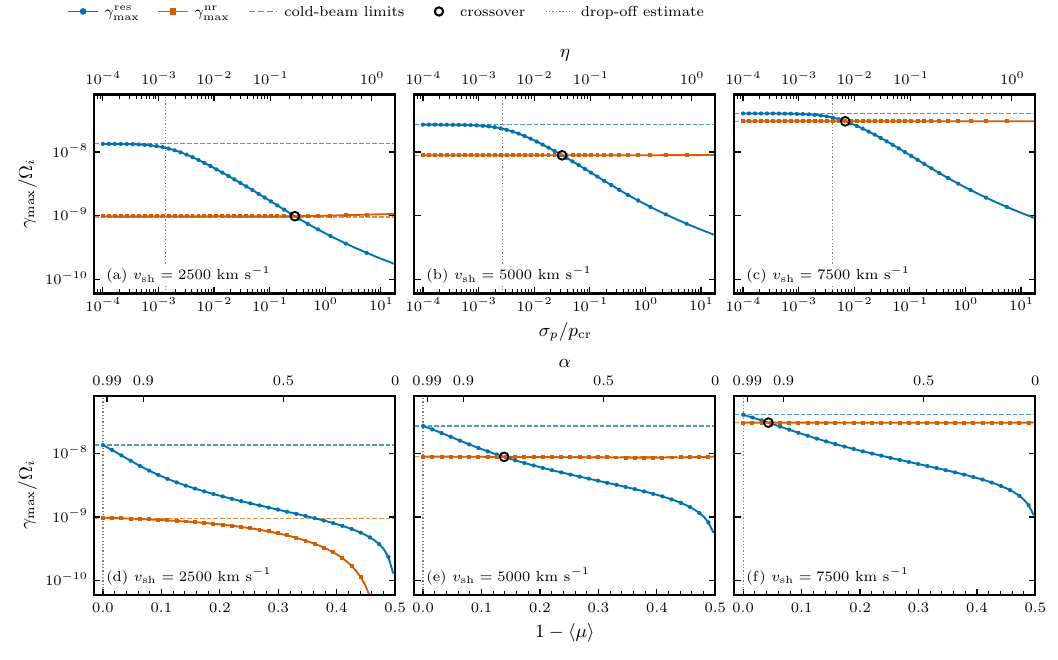}
\caption{\label{fig:growth-rate-vs-spread} Fastest-growing resonant (blue) and non-resonant (orange) growth rates as a function of momentum spread (top) and pitch-angle cosine spread (bottom) for three shock speeds (columns). Top row (a)--(c): Growth rates vs. momentum spread for a forward-beamed $\mu=1$ distribution. Bottom row (d)--(f): Growth rates vs. pitch-angle cosine spread for a monoenergetic $p=p_{cr}$ distribution. Horizontal lines mark the cold-beam estimates, Eqs.~\eqref{eq:gammafgm-non-res-full} and \eqref{eq:resonant-gammafgm-delta-approx}. Vertical lines denote the fastest-growing reactive resonant growth rate drop-off estimate as given by Eqs.~\eqref{eq:dropoff-estimate} and \eqref{eq:rms-equations-narrow}.}
\end{figure*}

\subsection{Numerical investigation} \label{sec:numerical-investigation}

To isolate the effects of momentum and pitch-angle spread, we perform two scans: varying $\sigma_p/p_{cr}$ at fixed $\mu=1$, and varying the anisotropy $1-\avg{\mu}$ at fixed $p=p_{cr}$. Since $\delta\mu=0$ in the former and $\delta p=0$ in the latter, the cross term in Eq.~\eqref{eq:dropoff-estimate} vanishes for both scans. Moreover, for a given shock speed, Eq.~\eqref{eq:Jcr-vshock-relation-general} fixes the corresponding CR current $J_{cr}=e n_{cr}\avg{v\mu}$. We therefore choose $n_{cr}$ so that this current remains fixed as either spread is increased beyond the cold-beam limit.

We take the CR distribution function to be
\begin{equation}\label{eq:cr-distribution-function}
    f_{cr}(p,\mu) =
    \frac{n_{cr}}{2\pi}
    \frac{g(p) h(\mu)}
    {\int_0^{\infty} g(p) p^2 \dd{p}\int_{-1}^1 h(\mu) \dd{\mu}} .
\end{equation}

The dependence on momentum magnitude is modeled as a log-normal distribution
\begin{equation} \label{eq:gp}
    g(p) = \pqty{\frac{p_{cr}}{p}}^2 \exp\qty{-\frac{\pqty{\ln p - \ln p_{cr}}^2}{2\eta^2}},
\end{equation}

\noindent where $\eta$ is the standard deviation of $\ln(p/p_{cr})$. The corresponding number spectrum $dN_{cr}/dp \propto p^2 g(p)$ is peaked at $p_{cr}$ and has been used as a model for escaping CRs \cite{Ohira+2010}. 

Pitch-angle cosine dependence is taken to be
\begin{equation} \label{eq:hmu}
    h(\mu) =
    \frac{\Theta(\mu)}{2}
    \frac{1-\alpha^2}
    {\pqty{1-2\mu\alpha + \alpha^2}^{3/2}},
\end{equation}

\noindent which is the Henyey--Greenstein function \cite{HenyeyGreenstein1941} restricted to $\mu>0$ by the Heaviside function $\Theta$. This choice is convenient because it is controlled by a single asymmetry parameter, $\alpha \in \lbrack 0,1)$; it reduces to an isotropic distribution over the forward hemisphere in the limit $\alpha \rightarrow 0$; and it approaches a forward beam in the limit $\alpha \rightarrow 1$. 

Restriction to $\mu>0$ reflects the defining property of the escaping CRs. Where scattering becomes inefficient, particles begin to stream ballistically away from the shock, with the returning CR flux small compared to the outgoing flux (Appendix~\ref{app:FEB-shock-relationship}). Distributions ranging from isotropic over the forward hemisphere to an ideal cold beam therefore span the physically appropriate range. Anisotropy carried by the bulk drift of a quasi-isotropic population instead characterizes the confined population near the shock (Sec.~\ref{sec:discussion}).

Although $\eta$ and $\alpha$ parameterize the distribution shapes, we label the scans by the more physical spreads $\sigma_p/p_{cr}$ and $1-\avg{\mu}$. From Eqs.~\eqref{eq:gp} and \eqref{eq:hmu}, they are related by

\begin{align}
    \sigma_p/p_{cr} &= \sqrt{e^{4\eta^2}-2e^{3\eta^2/2}+1}, \\
    \avg{\mu} &= \frac{\sqrt{1+\alpha^2}}{\sqrt{1+\alpha^2} + 1 - \alpha} \in \lbrack 0.5, 1).
\end{align}

Near the cold-beam limit, the rms deviations entering the analytic criterion in Eq.~\eqref{eq:dropoff-estimate} reduce to

\begin{equation} \label{eq:rms-equations-narrow}
    \sigma_p \simeq p_{cr} \eta \qc
    \sigma_{\mu} \simeq \sqrt{\frac{1-\avg{\mu}}{3}},
\end{equation}

\noindent as shown in Appendix~\ref{app:rms-p-mu-derivations}.

To obtain the growth rates of the fastest-growing resonant and non-resonant modes, we solve the dispersion relation Eq.~\eqref{eq:general-dispersion-relation} using Eq.~\eqref{eq:susceptibility} for the ion and electron susceptibilities. The CR susceptibility is computed numerically using Eq.~\eqref{eq:cr-distribution-function} in the gyrotropic susceptibility formula\footnote{Reference~\cite{AmatoBlasi2009} has the incorrect sign on its $\partial f_\alpha / \partial \mu$ term in Eq.~(1), which can be seen under the change of variables $\mu' = -\mu$.} \citep[see, e.g.,][]{Achterberg1983, AmatoBlasi2009}

\begin{equation} \label{eq:cr-susceptibility-general}
    \chi_{cr}^\sigma(\omega,k) = \frac{4\pi^2 e^2}{\omega}\int_0^\infty\!\dd{p}\int_{-1}^{1}\!\dd{\mu}\;\frac{p^2 v(1-\mu^2)}{a - b\mu}\,\mathcal{A}\bqty{f_{cr}},
\end{equation}

\noindent with $v(p) = p / \pqty\big{\Gamma(p) m_{cr}}$, $a(p) \equiv \omega + \sigma\Omega_{cr}/\Gamma(p)$, $b(p) \equiv k v(p)$, and

\begin{equation}
    \mathcal{A}\bqty{f_{cr}} \equiv \pdv{f_{cr}}{p} + \pqty{\frac{b}{\omega} - \mu}\frac{1}{p}\pdv{f_{cr}}{\mu}.
\end{equation}

The corresponding dispersion relation for a general gyrotropic CR distribution and its non-resonant unmagnetized limit are derived in Appendix~\ref{app:growth-rates-general}.

The derivative of the Heaviside in Eq.~\eqref{eq:hmu} supports an artificial high-wavenumber growing mode at $\mu=0$ which we suppress by replacing

\begin{equation} \label{eq:heaviside-smoothing}
    \Theta(\mu) \mapsto \frac{1}{2}\pqty{1 + \tanh\pqty{\mu/\epsilon}}
\end{equation}

\noindent for $\epsilon = 0.1$ when solving the dispersion relation numerically (see Fig.~\ref{fig:cr-distribution}(b)). 

\begin{figure}
\centering
\includegraphics[width=\columnwidth]{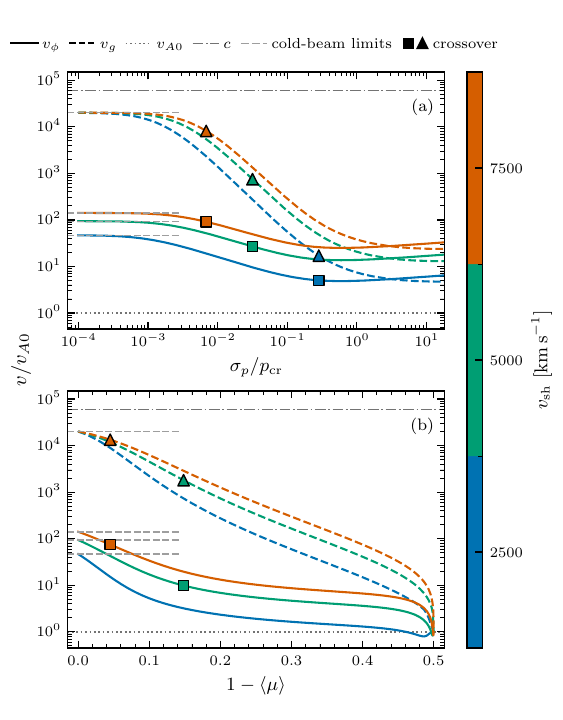}
\caption{\label{fig:phase-group-velocity-vs-spread}Phase velocity $v_\phi$ (solid) and group velocity $v_g$ (dashed) of the fastest-growing resonant mode as functions of momentum spread (a) and pitch-angle cosine spread (b). Squares (triangles) mark $v_{\phi}$ ($v_g$) at the crossover spread. Cold-beam limits are shown as truncated horizontal lines for readability. Subplot (b) uses the sharp-cutoff distribution ($\epsilon=0$) to follow the resonant branch toward the forward-hemisphere limit.}
\end{figure}

\begin{figure}
\centering
\includegraphics[width=\columnwidth]{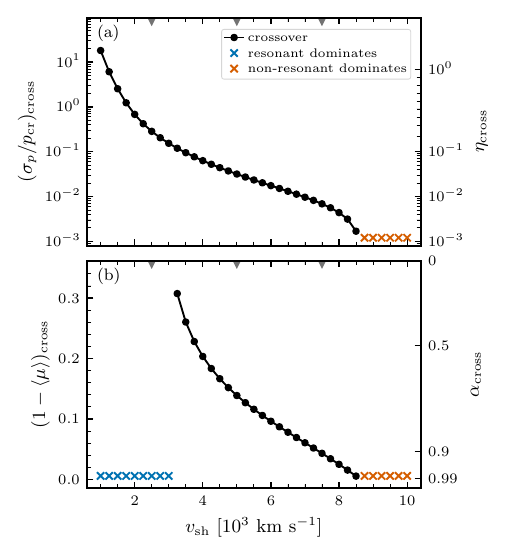}
\caption{\label{fig:crossover-vs-vshock}Spreads in momentum (a) and pitch-angle cosine (b) at which the resonant and non-resonant modes are equal (``crossover'' points), for shock speeds between $10^3\,{\rm km\,s^{-1}}$ and $10^4\,{\rm km\,s^{-1}}$ spaced every $250\,{\rm km\,s^{-1}}$. Black points denote shock speeds with a crossover; blue (orange) crosses mark shock speeds without one, where the resonant (non-resonant) mode dominates for all spreads considered. Gray triangles mark the three representative shock speeds used in Fig.~\ref{fig:growth-rate-vs-spread}.}
\end{figure}

Figure~\ref{fig:growth-rate-vs-spread}(a)--(c) shows how the resonant and non-resonant growth rates change from their cold-beam values as momentum-magnitude spread increases across three representative shock speeds. The resonant growth rate remains close to its cold-beam value until the drop-off estimate in each subplot, beyond which it steadily decreases. The non-resonant growth rate, by contrast, is essentially unaffected by momentum spread. The point at which the resonant and non-resonant growth rates are equal (the ``crossover'' point) occurs at larger spreads for slower shock speeds. 

Figure~\ref{fig:growth-rate-vs-spread}(d)--(f) similarly shows how the growth rates change with the anisotropy $1-\avg{\mu}$, with $0$ corresponding to a cold beam and $0.5$ an isotropic forward hemisphere. The resonant growth rate begins to decrease as soon as the mean pitch-angle cosine decreases. The crossover points again occur at larger pitch-angle cosine spreads for slower shocks, and interestingly, the non-resonant mode is unaffected by pitch-angle cosine spread except at sufficiently slow shocks, where it begins to stabilize at large spread. It is apparent from both scans that the drop-off estimate, Eq.~\eqref{eq:dropoff-estimate}, is conservative for the onset of decreased resonant growth rate. 

The spread in the escaping CR distribution has a significant effect on the resonant mode's Alfv\'enic character, as shown in Fig.~\ref{fig:phase-group-velocity-vs-spread}. Both subplots show an initial decrease from their cold-beam values once the distribution broadens. However, neither velocity approaches $v_{A0}$ over the range of momentum spreads considered [Fig.~\ref{fig:phase-group-velocity-vs-spread}(a)]. With increasing pitch-angle cosine spread [Fig.~\ref{fig:phase-group-velocity-vs-spread}(b)], both approach $v_{A0}$ only when the distribution is maximally broadened. Thus, the kinetic resonant instability may differ substantially from a conventional Alfv\'en-wave-like perturbation.

Subplots (a) and (b) of Fig.~\ref{fig:crossover-vs-vshock} show how the crossover points for the momentum- and pitch-angle-spread scans, respectively, vary with shock speed. Above $v_{\rm sh} \simeq 8500\,{\rm km\,s^{-1}}$, no crossover occurs in either subplot because the non-resonant mode dominates in the cold-beam limit, consistent with the crossover prediction of Eq.~\eqref{eq:cold-beam-crossover-vshock}. As seen in Fig.~\ref{fig:crossover-vs-vshock}(b), a crossover additionally does not occur at sufficiently low shock speeds because the non-resonant mode stabilizes with increasing $1-\avg{\mu}$. Such stabilization does not occur with increasing momentum spread over the shock speeds considered.

The broadness of escaping CR distributions at SNRs is poorly constrained by both theory and observation since it depends on the details of the transition between the diffusive and ballistic regimes near the free-escape boundary (FEB). In the literature, momentum distributions range from idealized monoenergetic distributions to finite-width distributions with $\sigma_p \sim p_{cr}$ \cite{Ohira+2010, Bell+2013, Zirakashvili+2008, Ellison+2011}, while pitch-angle distributions vary from almost hemispherical to forward-beamed \cite{Caprioli+2010c, Malkov+2011}. Despite this uncertainty, Fig.~\ref{fig:crossover-vs-vshock} shows that the resonant mode can dominate at sufficiently slow shocks even for broad CR distributions: the crossover point shifts to larger spreads as the shock speed decreases, and sufficiently broad distributions can stabilize the non-resonant mode while the resonant mode remains unstable.

\section{Hybrid Simulations} \label{sec:hybrid-pic-simulations}

\begin{figure}
\centering
\includegraphics[width=\columnwidth]{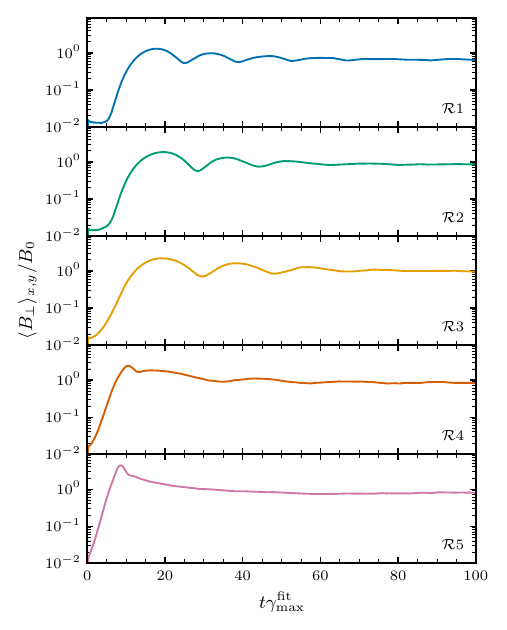}
\caption{\label{fig:bperp-evolution}Box-averaged perpendicular magnetic field over time for runs $\run{1}$--$\run{5}$, where time is normalized using the run's own fastest fitted growth rate, $\gamma_{\max}^{\rm fit} = \max(\gamma_{\max}^{\rm res, fit}, \gamma_{\max}^{\rm nr, fit})$.}
\end{figure}

\begin{figure}
\centering
\includegraphics[width=\columnwidth]{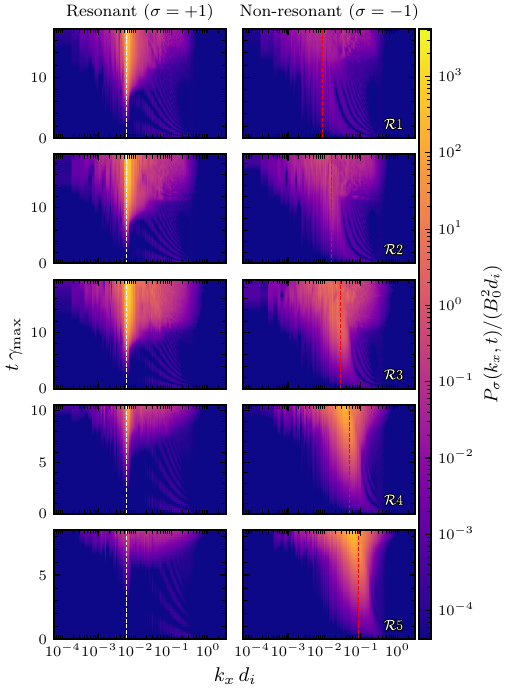}
\caption{\label{fig:power-spectra-evolution}Magnetic power spectra decomposed by helicity, $P_\sigma(k_x,t)$, as a function of the parallel wavenumber $k_x$ and time $t$ for runs $\run{1}$--$\run{5}$. The left column shows the resonant ($\sigma=+1$) spectrum with white lines indicating the predicted fastest-growing wavenumber from Eq.~\eqref{eq:resonant-kfgm-approx}. The right column shows the non-resonant ($\sigma=-1$) spectrum, with red lines indicating the predicted fastest-growing wavenumber from Eq.~\eqref{eq:non-resonant-kfgm}. Each subplot extends in time until the first maximum of $\avg{B_\perp}_{x,y}$.}
\end{figure}

\begin{figure}
\centering
\includegraphics[width=\columnwidth]{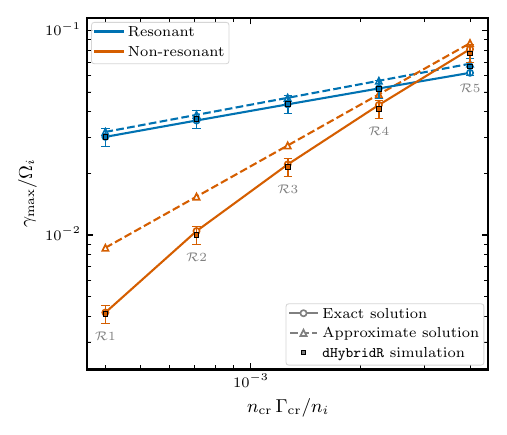}
\caption{\label{fig:growth-rate-sim-vs-theory}Growth rate of the fastest-growing resonant and non-resonant modes as a function of $n_{cr}\Gamma_{cr}/n_i$. Solid squares denote rates measured from runs $\run{1}$--$\run{5}$, solid lines with open circles denote the exact analytic solution to the cold-beam dispersion relation (Appendix~\ref{app:exact-dr-solution}), and dashed lines with open triangles denote the analytic approximations.}
\end{figure}

To test the fastest-growing-mode predictions of the previous sections, we perform five simulations using the relativistic hybrid particle-in-cell code \texttt{dHybridR} \cite{Haggerty+2019}. The code evolves ions and relativistic CRs kinetically under the influence of self-consistent electric and magnetic fields, with electrons acting as a charge-neutralizing, massless adiabatic fluid. All runs share common parameters except for the value of $n_{cr}/n_i$, which we logarithmically space between $2\times 10^{-4}$ and $2\times 10^{-3}$ (Table~\ref{tab:run-parameters}). These values are chosen so that our simulations sweep across the resonant-dominant (low $n_{cr}/n_i$) and non-resonant-dominant (large $n_{cr}/n_i$) regimes. 

\begin{table}
\centering
\begin{tabular}{lccc}
\toprule
Run & $n_{cr}/n_i$ & $\gamma_{\max}^{\rm res}/\Omega_i$ & $\gamma_{\max}^{\rm nr}/\Omega_i$ \\
\midrule
$\run{1}$ & $2.00\times 10^{-4}$ & $0.0302$ & $0.0042$ \\
$\run{2}$ & $3.56\times 10^{-4}$ & $0.0363$ & $0.0105$ \\
$\run{3}$ & $6.32\times 10^{-4}$ & $0.0435$ & $0.0221$ \\
$\run{4}$ & $1.12\times 10^{-3}$ & $0.0520$ & $0.0432$ \\
$\run{5}$ & $2.00\times 10^{-3}$ & $0.0620$ & $0.0810$ \\
\bottomrule
\end{tabular}
\caption{Values of $n_{cr}/n_i$ across the runs, with exact analytic predictions for the fastest-growing resonant and non-resonant cold-beam growth rates (Appendix~\ref{app:exact-dr-solution}) as shown in Fig.~\ref{fig:growth-rate-sim-vs-theory}.}
\label{tab:run-parameters}
\end{table}

Each simulation consists of a background thermal ion population with thermal speed $v_{{\rm th},i}\equiv \sqrt{k_{\rm B} T_i / m_i} = v_{A0}$, adiabatic fluid electrons with initial temperature $T_e \simeq T_i$, and a cold relativistic CR beam with $\Gamma_{cr} = 2$ propagating parallel to an external magnetic field, $\vb{B}_0 = B_0 \vu{x}$. The speed of light is taken to be $c = 100\,v_{A0}$. 

The simulations are two-dimensional in space and three-dimensional in velocity space (2D3V), with a box size $L_x \times L_y = 55000\; d_i \times 5 \; d_i$ and grid spacing $\Delta x = \Delta y = 1\; d_i$. Since the predicted fastest-growing resonant wavelength $\lambda_{\rm res} = 2\pi / k_{\max}^{\rm res} = 2\pi r_{cr}$ is the same in every run, we capture roughly $50$ of these wavelengths along the $x$ axis across the suite. The small $L_y$ implies the simulations are spatially quasi-1D. 

The electromagnetic fields and background ions have periodic boundary conditions along all axes. The CRs are periodic along $y$, open along $x$, and continuously injected into the box at $x=0$. Ions and CRs are initialized uniformly throughout the box, each with 64 ($8\times 8$) macroparticles per cell. The timestep is chosen to satisfy the CFL condition, $\Delta t = C_{\max} / \pqty{c\sqrt{1/(\Delta x)^2 + 1/(\Delta y)^2}}$ for safety factor $C_{\max} = 0.5$. 

For each run, we track magnetic field amplification by computing the box-averaged perpendicular magnetic field magnitude $\avg{B_\perp}_{x,y}$ where $B_\perp = \sqrt{B_y^2 + B_z^2}$ (Fig.~\ref{fig:bperp-evolution}). We assess which modes contribute most to this amplification through the $y$-averaged power spectrum,

\begin{equation}
    P_\sigma(k_x,t) \equiv \frac{1}{L_x} \avg{\abs{\widehat{B}_\sigma(k_x,y,t)}^2}_y,
\end{equation}

\noindent where $B_\sigma = \vb{B}_\perp \vdot \vu{e}_\sigma^* = (B_y - i \sigma B_z) / \sqrt{2}$ and 

\begin{equation}
    \widehat{B}_\sigma(k_x,y,t) = \frac{1}{\sqrt{2\pi}}\int_0^{L_x} B_\sigma(x,y,t)e^{-ik_xx}\dd{x}
\end{equation}

\noindent denotes the Fourier transform along $x$.

\subsection{Linear growth}

Figure~\ref{fig:bperp-evolution} shows that the perpendicular magnetic field initially grows exponentially in every run, with larger values of $n_{cr}/n_i$ producing faster growth. Figure~\ref{fig:power-spectra-evolution} identifies which mode drives this growth across runs. Runs $\run{1}$--$\run{3}$ have the most growth in a narrow band around the predicted fastest-growing resonant wavenumber (Eq.~\eqref{eq:resonant-kfgm-approx}), which is common to all the runs since $r_{cr}$ is fixed. Run $\run{4}$ shows a roughly equal competition between the narrow-band resonant and broader-band non-resonant modes, with the non-resonant mode clearly dominating in $\run{5}$.

To quantify this mode competition for each run, we average each power spectrum over a chosen band of wavenumbers $k_x$ around the respective predicted fastest-growing wavenumber and then fit $\ln \avg{P_\sigma}_{k_x} = \text{const} + 2\gamma t$ over time, allowing us to extract an approximate simulation growth rate for each mode. The time intervals for the fits were chosen to be well into the linear regime---but before saturation---where $\ln \avg{P_\sigma}_{k_x}$ appears linear. We assign each simulation growth rate a $10\%$ uncertainty to account for variations in the fitted growth rate due to the chosen $k_x$ band, time interval, and inherent power spectrum noise. 

Extracting growth rates in this manner is more accurate than fitting the perpendicular magnetic field directly since the latter method's growth rate matches a particular mode only when that mode dominates. Because the initial magnetic field noise also varies with wavenumber, the perpendicular field may not reflect the growth rate of the fastest-growing mode if that mode starts from a low noise level.

Figure~\ref{fig:growth-rate-sim-vs-theory} shows these simulation-extracted growth rates as a function of $n_{cr} \Gamma_{cr} / n_i$. For comparison, we plot the exact analytic cold-beam growth rates (Appendix~\ref{app:exact-dr-solution}), and the approximate growth rates for the resonant (Eq.~\eqref{eq:resonant-gammafgm-delta-approx}) and non-resonant (Eq.~\eqref{eq:non-resonant-gammafgm-approx}) modes. The exact cold-beam growth rates agree closely with the simulations, while the approximate non-resonant formula becomes a poor predictor as $n_{cr} \Gamma_{cr} / n_i$ decreases. This is expected:  Eq.~\eqref{eq:non-resonant-gammafgm-approx} assumes $n_{cr} \Gamma_{cr} / n_i \gg (v_{A0} / v_{cr})^2$, a condition that is increasingly violated as the CRs become more dilute relative to the background ions. In contrast, the approximate formula for the resonant instability remains accurate within $\sim10\%$ since it requires the less restrictive condition $n_{cr} \Gamma_{cr} / n_i \gg (v_{A0} / v_{cr})^3$. 

\subsection{Nonlinear evolution and saturation}

\begin{figure}
\centering
\includegraphics[width=\columnwidth]{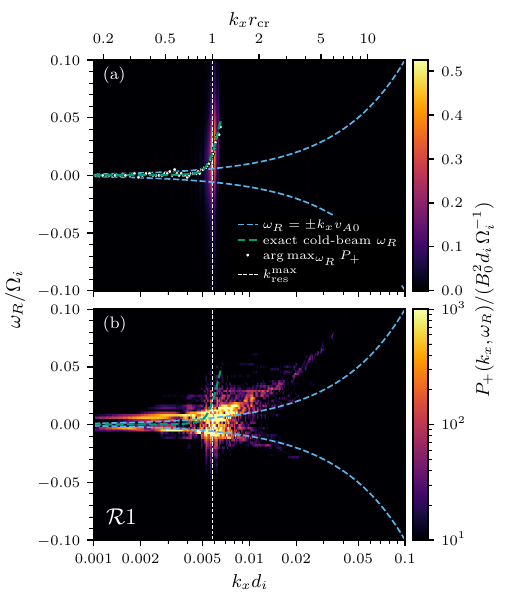}
\caption{\label{fig:k-omega-spectrum-linear-vs-saturation}Magnetic power spectrum of the resonant mode, $P_+(k_x,\omega_R)$, as a function of the parallel wavenumber $k_x$ and real frequency $\omega_R$ for run $\run1$. Subplot (a) shows power during the linear regime, $t\gamma_{\max}^{\rm fit} \in \bqty{2.5,6}$, while subplot (b) shows power well into saturation, $t\gamma_{\max}^{\rm fit} \in \bqty{60, 150}$. Blue dashed lines denote forward- and backward-propagating Alfv\'en waves, $\omega_R = \pm k_x v_{A0}$. The green dashed curve denotes Eq.~\eqref{eq:wR-exact} where $\mathcal{D}>0$. In (a), white points denote the frequencies at which $P_+$ is maximized at each $k_x$.}
\end{figure}

After the linear stage, the perpendicular magnetic field in every run reaches a maximum and then decays toward a lower saturation value around $\avg{B_\perp}_{x,y} \sim B_0$ (Fig.~\ref{fig:bperp-evolution}), with larger values of $n_{cr}/n_i$ producing an earlier, higher overshoot. Runs $\run{1}$--$\run{3}$ show oscillations similar to those observed for cold ring distributions \cite{Lemmerz+2025}, a signature of the reactive resonant instability. These oscillations damp over time, weaken in $\run4$, and cease in $\run5$ once the non-resonant mode becomes dominant.

To illustrate how the resonant mode changes from reactive to quasi-Alfv\'enic as the magnetic field saturates, we resolve the power spectrum in wavenumber and frequency,

\begin{equation} \label{eq:komega-power-spectrum}
    P_\sigma(k_x,\omega_R) \equiv \frac{1}{L_x T}\avg{\abs{\widehat{B}_\sigma(k_x,y,\omega_R)}^2}_y,
\end{equation}

\noindent where $T$ is the duration of the time window and

\begin{equation}
    \widehat{B}_\sigma(k_x,y,\omega_R) = \frac{1}{\sqrt{2\pi}}\int_{t_0}^{t_0+T} \widehat{B}_\sigma(k_x,y,t)\,e^{i\omega_R t}\dd{t}
\end{equation}

\noindent is the Fourier transform in time. Figure~\ref{fig:k-omega-spectrum-linear-vs-saturation}(a) and (b) show the resonant spectrum for run $\run1$ during the linear regime and after saturation, respectively (runs $\run2$ and $\run3$, not shown, are similar). In the linear regime, the frequency at which the power spectrum is maximized at each $k_x$ closely matches the exact cold-beam resonant frequency, Eq.~\eqref{eq:wR-exact}, and is consistent with the steep slope at $k_{\max}^{\rm res}$ set by the group velocity of the fastest-growing resonant mode, Eq.~\eqref{eq:fgm-group-velocity}. After saturation, the power moves toward $\omega_R \simeq \pm k_x v_{A0}$, implying the fluctuations are more Alfv\'enic than reactive. Their presence at both positive and negative $\omega_R$ demonstrates propagation parallel and antiparallel to $\vb{B}_0$. 

Once runs $\run1$--$\run3$ have saturated at late times, around $45\%$ of the initially $\mu=1$ CRs have obtained $\mu<0$, demonstrating order unity pitch-angle redistribution. Moreover, defining the rms magnetic field around the resonant scale,

\begin{equation}
    \delta B_{\rm rms}^{\rm res}(t) \equiv \bqty{ 2\int_{1/(2r_{cr})}^{2/r_{cr}} P_+(k_x,t) \dd{k_x} }^{1/2},
\end{equation}

\noindent and time-averaging over the last $100\,\Omega_i^{-1}$ gives $\avg{\delta B_{\rm rms}^{\rm res}}_t / B_0 \in \qty{0.54, 0.67, 0.76}$ for $\run{1}$--$\run{3}$, respectively. The standard parallel mean free path estimate, $\lambda_\parallel / r_{cr} \sim (B_0 / \delta B_{\rm rms}^{\rm res})^2$, then yields mean free paths on the order of $r_{cr}$. Together with the order-unity pitch-angle redistribution, this indicates that saturated magnetic fluctuations efficiently scatter the escaping CRs within the simulation domain. A detailed study of CR scattering at saturation is deferred to future work.

\section{Discussion} \label{sec:discussion}

The reactive resonant and non-resonant modes of Sec.~\ref{sec:general-dispersion-relation} have a long lineage in the space-plasma literature, where they are known as the ion/ion right-hand resonant and ion/ion non-resonant instabilities \cite{WinskeGary1986, Gary1993}, with ``right-handed'' referring to the polarization $\mathcal{P}=+1$ of the resonant mode in the ion bulk-rest frame (Table~\ref{tab:sigma-helicity-polarization}). Our results generalize the reactive resonant instability to relativistic beams, which carry an additional factor of $\Gamma_{cr}^{-2/3}$ in their growth rate (Eq.~\eqref{eq:resonant-gammafgm-delta-approx}). We additionally provide exact, closed-form solutions for both cold-beam instabilities and discuss their connection under parity transformation (Appendix~\ref{app:exact-dr-solution}). 

The competition between the resonant and non-resonant modes was previously explored by Haggerty et al. \cite{Haggerty+2019p}, who measured growth rate curves and helicities for power-law and forward-beamed CR distributions using quasi-1D, undriven \texttt{dHybridR} simulations, deferring the underlying linear theory to a forthcoming paper. The present work supplies that theory and driven simulations in the context of an escaping CR population at SNR shocks.

Our analysis parallels the calculations of Bell \cite{Bell2004} and Amato and Blasi \cite{AmatoBlasi2009}, who compared the resonant and non-resonant modes for the CR population confined at the shock. In the shock frame, their CR distribution follows a stationary, isotropic power
law whose number density is dominated by particles at its minimum momentum cutoff $p_{\rm min}$: the injection momentum $p_{\rm inj}$ in Amato and Blasi's analysis and a local minimum momentum in Bell's. In the ion bulk-rest frame, it therefore drifts at $v_{\rm sh}$ with all anisotropy attributable to this drift rather than to angular structure in the distribution itself. The escaping population considered here is the opposite limit in both respects: it is at most isotropic over the forward hemisphere, with anisotropy specified by the distribution itself, and its number density is peaked at the escape momentum $p_{cr} = p_{\max}$.

In the works cited above, the resonant growth rate peaks at $k r_g(p_{\rm min}) \sim 1$ for gyroradius $r_g(p) = p c/(eB_0)$ while ours peaks at $k r_g(p_{\max}) \sim 1$: the amplified fluctuations are thus on the scale required to scatter the highest-energy CRs rather than far below it. Moreover, a population with sufficiently small pitch-angle and momentum spreads may respond coherently, producing a reactive resonant instability. This reactive solution is excluded by construction in the standard treatment \cite{Bell2004, AmatoBlasi2009}, which neglects $\omega$ in the resonant denominator. Such an omission is valid for a broad, isotropic distribution, but discards the reactive root of a beamed escaping population.

Amato and Blasi's large-$k$ estimate for the disappearance of the non-resonant mode (their Eq.~(37)) yields a shock velocity of $\sim 600 \,{\rm km\,s^{-1}}$ at our fiducial parameters, with their non-resonant mode dominating whenever it exists. The situation is quite different for the escaping population far upstream: the reactive resonant mode dominates at shock speeds up to $8600 \,{\rm km\,s^{-1}}$ at our fiducials (Table~\ref{tab:fiducial-snr-parameters}), with broadened distributions inducing a kinetic resonant instability that remains dominant for shock speeds $\lesssim 2000 \,{\rm km\,s^{-1}}$.

As is known in the literature, the non-resonant mode requires a sufficiently strong current to become unstable. However, as Fig.~\ref{fig:growth-rate-vs-spread}(d) demonstrates, the non-resonant mode may stabilize under pitch-angle broadening even for a fixed CR current. This occurs because broadening moves CRs to small parallel velocities where they begin to magnetize, $\abs{\omega''} = \Gamma\abs{\omega - kv_\parallel} \lesssim \Omega_{cr}$, regardless of their gyroradii. The collective response of these particles---the last term in Eq.~\eqref{eq:dispersion-relation-factored-general}---then becomes large enough to counteract the unmagnetized CR current driving the non-resonant mode. This is distinct from stabilization at low CR current and, to our knowledge, has not been reported before.

While we have focused on CRs escaping SNR shocks, similar considerations apply to the self-confinement regions that CRs establish around their sources \cite{Malkov+2013, Blasi+2015, Evoli+2018, Nava+2016, Nava+2019, Schroer+2021, Cermenati+2026}. For CRs to escape such a region, scattering must eventually become inefficient with increasing distance from the source, leading to an increasingly anisotropic CR population near the region's boundary. The resonant and non-resonant growth rates induced by such a population can differ significantly from those of a quasi-isotropic, drifting distribution. These results may therefore affect estimates of both the extent of self-confinement regions and the flux of CRs escaping into the surrounding medium.

These results also bear on the maximum attainable CR energy at SNR shocks. The value of $E_{\max}$ is commonly obtained by requiring the escaping CR current to amplify magnetic perturbations via the non-resonant instability by several e-folds in a parcel of upstream fluid before the shock overtakes it \cite{Bell+2013}. In situations where the resonant mode dominates, its growth at $k r_g(p_{\max}) \sim 1$ and alternative scaling with $n_{cr}/n_i$ may modify the resulting value of $E_{\max}$, with potential implications for whether SNRs accelerate CRs to the knee of the CR energy spectrum. A quantitative study of $E_{\max}$ in the resonant-dominant regime is deferred to future work.

\section{Conclusion}

Motivated by the beam-like character of CRs escaping far upstream of collisionless shocks, we obtained both approximate and exact solutions of the corresponding reactive resonant and non-resonant dispersion relations. We then quantified how finite pitch-angle and momentum spreads modify the growth rates at fixed CR current, with particular focus on the transition of the resonant mode from reactive to kinetic. Although the resonant growth rate falls by up to two orders of magnitude across this transition, it can remain dominant over the non-resonant mode even when the system is strongly current-driven. The non-resonant growth rate, by contrast, is largely insensitive to the shape of the CR distribution when the CRs are unmagnetized; sufficient pitch-angle broadening, however, can magnetize enough of the CRs to extinguish the mode entirely.

Hybrid particle-in-cell simulations spanning the transition between the two instabilities confirm the cold-beam linear theory. In the resonant-dominated runs, the reactive mode of the linear stage becomes more Alfv\'enic at saturation with power near $\omega_R \simeq \pm k v_{A0}$. These saturated fluctuations reach $\delta B/B_0 \sim 1$, produce order-unity pitch-angle redistribution, and yield estimated mean free paths comparable to the driving CR gyroradius. The resonant instability can therefore amplify the far-upstream field to $\delta B/B_0 \sim 1$ and strongly scatter the escaping CRs.

The resulting picture qualifies the standard expectation that the non-resonant mode dominates whenever the system is strongly current-driven (Appendix~\ref{app:stability-non-res}). Far upstream, the escaping CRs induce a resonant mode that dominates below a distribution-dependent crossover shock speed of order $10^3$ to $10^4\,{\rm km\,s^{-1}}$, with broader distributions corresponding to smaller crossover speeds (Fig.~\ref{fig:crossover-vs-vshock}). This crossover speed far exceeds that of a drifting, quasi-isotropic power law confined at the shock \cite{Bell2004, AmatoBlasi2009}, a difference attributable entirely to the escaping distribution's anisotropy and concentration at the maximum momentum. Calculations that ignore the resonant instability systematically miss its fastest-growing mode's contribution to amplification at the very scale needed to efficiently scatter escaping CRs. 

\begin{acknowledgments}
We thank P.~Blasi for stimulating discussions on CR-driven instabilities. Simulations were performed on computational resources provided by the University of Chicago Research Computing Center. D.C. was partially supported by NASA grant 80NSSC18K1726 and NSF grant AST-2510951.
\end{acknowledgments}

\appendix

\section{Circular basis sign, polarization, and helicity} \label{app:basis-sign-pol-hel}

A general transverse magnetic fluctuation may be written in the Cartesian and circular bases as

\begin{equation} \label{eq:dBperp-general}
    \vb{B}_\perp = B_y\vu{y} + B_z\vu{z} = B_\sigma \vu{e}_\sigma + B_{-\sigma} \vu{e}_{-\sigma},
\end{equation}

\noindent where $\vu{e}_\sigma = (\vu{y}+i\sigma\vu{z})/\sqrt{2}$ are the circular basis vectors for $\sigma = \pm 1$ and $\vu{e}_\sigma \vdot \vu{e}_{\sigma'}^* = \delta_{\sigma\sigma'}$. The reality of $\vb{B}_\perp$ (namely, $\vb{B}_\perp = \vb{B}_\perp^*$) provides a constraint on the circular basis coefficients, $B_\sigma = B_{-\sigma}^*$. 

The fluctuation is circularly polarized when $B_\sigma = C_\sigma e^{i(\phi+\alpha_\sigma)}$ for constant real amplitude $C_\sigma \geq 0$, phase $\phi = k_x x-\omega_R t$, and phase offset $\alpha_\sigma$. We take $\alpha_\sigma = 0$ for simplicity. Given this, Eq.~\eqref{eq:dBperp-general} becomes

\begin{align}
    \vb{B}_\perp &= C_\sigma e^{i\phi} \vu{e}_\sigma + C_\sigma e^{-i\phi}\vu{e}_{-\sigma} \label{eq:dBperp-circular}\\
                 &\propto \vu{y}\cos\phi - \sigma \vu{z} \sin\phi \label{eq:dBperp-final}.
\end{align}

By \textit{left-} or \textit{right-hand polarization} of a circularly polarized mode, we mean the rotation of Eq.~\eqref{eq:dBperp-final} in time at a fixed point in space (e.g., $x=0$) as we look along $\vb{B}_0 \parallel +x$. Concretely, this will be the rotation of the vector

\begin{equation}
    \vb{B}_\perp \propto \vu{y}\cos\pqty{\omega_R t} + \sigma \vu{z} \sin\pqty{\omega_R t}.
\end{equation}

Polarization can be written succinctly as $\mathcal{P} = \sigma \sgn (\omega_R)$, where $\mathcal{P} = +1$ corresponds to \textit{right-handed} rotation in time and $\mathcal{P} = -1$ \textit{left-handed} rotation. 

By \textit{left-} or \textit{right-hand helicity} of a circularly polarized mode, we mean the rotation of Eq.~\eqref{eq:dBperp-final} in space at a fixed moment in time (e.g., $t=0$). Concretely,

\begin{equation}
    \vb{B}_\perp \propto \vu{y}\cos\pqty{k_x x} - \sigma \vu{z} \sin\pqty{k_x x}.
\end{equation}

Defining $\mathcal{H} = \sigma \sgn (k_x)$, $\mathcal{H} = +1$ corresponds to \textit{left-handed} rotation in space and $\mathcal{H} = -1$ \textit{right-handed} rotation\footnote{Equivalently, $\mathcal H$ is the eigenvalue of $i\vu{k}\cross$ acting on $\vu{e}_\sigma$: $i\vu{k}\cross\vu{e}_\sigma=\mathcal H\vu{e}_\sigma$.}. This definition is identical to dimensionless magnetic helicity as given in \cite{Gary1993} for a parallel-propagating circularly polarized mode. Unlike polarization, the mode's handedness label is identical when viewed along $+x$ or $-x$.

Importantly, if $\omega_R$ and $k_x$ have the same sign, then $\mathcal P=\mathcal H$; however, because positive $\mathcal P$ denotes right-handed rotation in time whereas positive $\mathcal H$ denotes left-handed rotation in space, the corresponding handedness labels are opposite.

\begin{table}[t]
\centering
\begin{tabular}{ccc cc}
\toprule
 & & & \multicolumn{2}{c}{\textit{implies}} \\
\cmidrule(l){4-5}
$\sigma$ & $\sgn (k_x)$ & $\sgn (\omega_R)$ & helicity & polarization \\
\midrule
\multirow{4}{*}{$+1$}
 & $+$ & $+$ & LH & RH \\
 & $+$ & $-$ & LH & LH \\
 & $-$ & $+$ & RH & RH \\
 & $-$ & $-$ & RH & LH \\
\midrule
\multirow{4}{*}{$-1$}
 & $+$ & $+$ & RH & LH \\
 & $+$ & $-$ & RH & RH \\
 & $-$ & $+$ & LH & LH \\
 & $-$ & $-$ & LH & RH \\
\bottomrule
\end{tabular}
\caption{Given the signs of $\sigma$, $k_x$, and $\omega_R$, the helicity and polarization are uniquely determined as given above.}
\label{tab:sigma-helicity-polarization}
\end{table}

Table~\ref{tab:sigma-helicity-polarization} summarizes how helicity and polarization depend on the signs of $\sigma$, $k_x$, and $\omega_R$. Throughout this paper we take $k_x>0$ while allowing $\omega_R$ and $\sigma$ to have either sign. With this choice, the helicity is fixed directly by the circular basis sign, $\mathcal{H}=\sigma$, and only the $k_x>0$ rows of the table are used. Other sign conventions are possible: for example, one could instead fix $\sigma=+1$ and allow $k_x$ and $\omega_R$ to be positive or negative, in which case the top block would be relevant. 

Let us also consider the frame-dependence of polarization and helicity. Boosting along $v_x > 0$ transforms the real frequency and wavenumber via
\begin{align}
    k_x' &= \Gamma_v \pqty{k_x - v_x\omega_R / c^2}, \\
    \omega_R' &= \Gamma_v \pqty{\omega_R - v_x k_x}.
\end{align}

In the limit of non-relativistic boosts, helicity is invariant but polarization is not; polarization flips whenever $v_x > v_\phi > 0$, a condition routinely met in practice. Helicity is likewise not invariant under general Lorentz boosts, flipping for superluminal phase velocities satisfying $v_\phi > c^2 / v_x$. Since this latter condition is much harder to achieve, helicity is the more robust classification.

If a sufficiently large Lorentz boost gives $k'_x< 0$, we maintain our positive-wavenumber convention by recognizing that from Eq.~\eqref{eq:dBperp-circular}, 

\begin{equation} \label{eq:Bperp-parity}
    \vb{B}_\perp (\phi, \sigma) = \vb{B}_\perp (-\phi,-\sigma).
\end{equation}

Therefore, after boosting, we can perform the relabeling $(\omega_R, k_x, \sigma) \mapsto (-\omega_R, -k_x, -\sigma)$ \citep[cf.][App.~B]{Shalaby+2021}. This change is mere bookkeeping: both polarization $\mathcal{P}$ and helicity $\mathcal{H}$ in the boosted frame are unchanged\footnote{More generally, for a mode with $\omega=\omega_R+i\gamma$ and $k_x=k_{x,R}+i\kappa_x$, a Lorentz boost mixes the temporal and spatial growth/decay rates: $\gamma'=\Gamma_v(\gamma-v_x\kappa_x)$ and $\kappa_x'=\Gamma_v(\kappa_x-v_x\gamma/c^2)$. The corresponding relabeling for the full complex mode is $(\omega',k_x',\sigma)\mapsto (-\omega'^{*},-k_x'^{*},-\sigma)$.}. 

\section{Exact solution to the cold dispersion relation} \label{app:exact-dr-solution}

For ease of notation in what follows, we define the dimensionless variables $\tw = \omega / \Omega_i, \tgamma = \gamma / \Omega_i, \tk = kd_i, \tvcr = v_{cr} / v_{A0}$, and $\tncr = n_{cr}/n_i$. 

Clearing the resonant denominator in Eq.~\eqref{eq:dispersion-relation-factored} yields the cubic

\begin{equation}\label{eq:cubic-w}
    \tw^3 + a_2\tw^2 + a_1 \tw + a_0 = 0,
\end{equation}

\noindent with coefficients

\begin{align}
    a_2 &= \sigma\pqty{\tncr + \frac{1}{\Gamma_{cr}}} - \tk\tvcr, \\
    a_1 &= -2\sigma \tk \tncr \tvcr - \tk^2,\\
    a_0 &= \tk^2\pqty{\tk\tvcr + \sigma\pqty{\tncr \tvcr^2 - \frac{1}{\Gamma_{cr}}}}.
\end{align}

Substituting $\tw = \twR + i\tgamma$ into Eq.~\eqref{eq:cubic-w} and taking the imaginary part yields 

\begin{equation}\label{eq:gammasq-exact}
    \tgamma^2 = 3\twR^2 + 2a_2\twR + a_1
\end{equation}

\noindent for $\tgamma \neq 0$. Then, substituting this into the real part of the same formula produces a cubic for the real frequency,

\begin{equation}\label{eq:cubic-wR}
    \twR^3 + a_2 \twR^2 + \frac{a_1 + a_2^2}{4} \twR + \frac{a_1 a_2 - a_0}{8} = 0.
\end{equation}

The solution to Eq.~\eqref{eq:cubic-wR} is readily obtained via Cardano's formula; specifically, 

\begin{equation} \label{eq:wR-exact}
    \twR(k) = \sqrt[3]{-\frac{\mathcal{S}}{2} + \sqrt{\mathcal{D}}} + \sqrt[3]{-\frac{\mathcal{S}}{2} - \sqrt{\mathcal{D}}} - \frac{a_2}{3},
\end{equation}

\begin{align}
    \mathcal{D} &= \pqty{\frac{\mathcal{S}}{2}}^2 + \pqty{\frac{\mathcal{Q}}{3}}^3, \label{eq:eqD} \\
    \mathcal{Q} &= \frac{3a_1-a_2^2}{12}, \\
    \mathcal{S} &= \frac{9a_1 a_2 -2a_2^3 - 27a_0}{216},
\end{align}

\noindent where $\sqrt[3]{\cdot}$ is the real---not principal---cube root and $\mathcal{D}>0$ for growing modes. The growth rate $\tgamma(k)$ then equals the square root of Eq.~\eqref{eq:gammasq-exact}.

The exact solution makes manifest a relationship between the resonant and non-resonant branches under a reversal of the CR drift velocity: the $\sigma$ branch at $-v_{cr}$ has the same growth rate as the $-\sigma$ branch at $v_{cr}$, with the sign of the real frequency reversed,

\begin{align} \label{eq:parity-relation1}
    \gamma^\sigma(k;v_{cr}) &= \gamma^{-\sigma}(k;-v_{cr}),\\
    \omega_R^\sigma (k;v_{cr}) &= -\omega_R^{-\sigma}(k;-v_{cr}). \label{eq:parity-relation2}
\end{align}

Note that this mapping connects two physically distinct systems: one with $v_{cr}$, the other with $-v_{cr}$. 

This relationship holds beyond the cold-beam limit and follows from
parity invariance: the magnetic field $\vb{B}$---a pseudovector---is
unchanged under parity transformation, whereas regular vectors reverse
sign. Specifically, under the parity transformation
$\vb{x} \mapsto \vb{x}' = -\vb{x}$,

\begin{align}
    (\omega_R, \gamma, v_{cr}, k_x, \sigma) &\mapsto (\omega_R',\gamma',v_{cr}', k_x', \sigma') \nonumber\\
    &= (\omega_R, \gamma, -v_{cr}, -k_x, \sigma).
\end{align}

Since $k_x' = -k_x < 0$, we maintain the positive-wavenumber
convention by employing the relabeling from Eq.~\eqref{eq:Bperp-parity},
namely

\begin{align}
    (\omega_R',\gamma',v_{cr}', k_x', \sigma') &\mapsto (-\omega_R', \gamma', v_{cr}', -k_x', -\sigma') \nonumber\\
    &= (-\omega_R, \gamma, -v_{cr}, k_x, -\sigma).
\end{align}

Hence, the full mapping is $(\omega_R,\gamma,v_{cr}, k_x, \sigma) \mapsto
(-\omega_R, \gamma, -v_{cr}, k_x, -\sigma)$, matching
Eqs.~\eqref{eq:parity-relation1} and~\eqref{eq:parity-relation2}.

\section{Stability criterion for the non-resonant mode} \label{app:stability-non-res}

The non-resonant mode becomes marginally stable once the two complex-conjugate roots of Eq.~\eqref{eq:cubic-w} merge into a double real root, which occurs when $\mathcal{D}(k)=0$ \cite{Melrose1986}. Since the fastest-growing mode moves toward $k = 0$ as this condition is approached, the long-wavelength limit of $\mathcal{D}$ sets the criterion for stability. Using
Eq.~\eqref{eq:eqD}, we obtain

\begin{equation}
    \qty{\frac{\Gamma_{cr} n_{cr}}{n_i}\pqty{\frac{v_{cr}^2}{v_{A0}^2} -1} - 1} \pqty{k d_i}^2 + \order{(kd_i)^3} = 0,
\end{equation}

\noindent whose leading coefficient vanishes when

\begin{equation}
   \frac{\Gamma_{cr} n_{cr}}{n_i}\pqty{\frac{v_{cr}^2}{v_{A0}^2} -1} = 1.
\end{equation}

We note that this condition is derived for a cold beam; in general, its form is distribution-dependent.

The mode grows when the left-hand side exceeds unity and is stable at all wavenumbers otherwise. For $v_{cr} \gg v_{A0}$, this can be recast in two ways. The first,

\begin{equation}
    k_{\max}^{\rm nr} r_{cr} = \frac{1}{2},
\end{equation}

\noindent follows from Eq.~\eqref{eq:non-resonant-kfgm} and shows that the unmagnetized approximation breaks down at marginal stability: the wavenumber it predicts violates the $k r_{cr} \gg 1$ assumption used to derive it. The stability criterion, then, cannot be obtained from Eq.~\eqref{eq:non-resonant-gammak}; zeroing the growth rate of its fastest-growing mode, Eq.~\eqref{eq:gammafgm-non-res-full}, gives a threshold $4\times$ too large.

The second recasting is

\begin{equation} \label{eq:current-eq}
    \frac{4\pi J_{cr} r_g(p_{\max})}{cB_0} = 1
\end{equation}

\noindent using $r_g(p) = p c/(eB_0)$ and $r_g(p_{\max}) = r_{cr}$. Amato and Blasi \cite{AmatoBlasi2009} found an analogous stability criterion within a factor of order unity, where their gyroradius is given in terms of the injection momentum, $r_g(p_{\rm inj})$. Following their convention, a system is strongly
current-driven when $4\pi J_{cr} r_g/(cB_0) \gg 1$. 

\section{Derivation of Eq.~\eqref{eq:rms-equations-narrow}} \label{app:rms-p-mu-derivations}

The mean-squared deviation from the cold-beam momentum, $p=p_{cr}$, is given in terms of $\eta$ via

\begin{equation}
    \sigma_p^2 \equiv \avg{\pqty{p-p_{cr}}^2} =  p_{cr}^2 \pqty{e^{4\eta^2}-2e^{3\eta^2/2}+1},
\end{equation}

\noindent which Taylor-expanded about $\eta=0$ gives $\sigma_p \simeq p_{cr}\eta$ accurate to $\order{\eta^3}$. 

The mean-squared deviation from the cold-beam pitch-angle cosine,
$\mu=1$, is defined as
\begin{equation}
    \sigma_\mu^2 \equiv \avg{(\mu-1)^2}
    =
    \frac{\int_{-1}^1 (1-\mu)^2 h(\mu) \dd{\mu}}
         {\int_{-1}^1 h(\mu) \dd{\mu}} .
\end{equation}
We begin by making the change of variables $\beta = 1-\alpha$ and
$\nu = 1-\mu$ so that values close to the cold-beam limit are small, giving
\begin{equation}
    h(\nu)
    =
    \frac{\Theta(1-\nu)}{2}
    \frac{1-(1-\beta)^2}
         {\pqty\big{\beta^2 + 2(1-\beta)\nu}^{3/2}} .
\end{equation}

Then, for $\beta^2 \ll \nu \leq 1$,

\begin{equation}
    h(\nu) \propto \Theta(1-\nu) \nu^{-3/2}.
\end{equation}

Using this proportionality over $\nu\in\bqty{0,1}$, we obtain to leading order

\begin{equation}
    \frac{\avg{(1-\mu)^2}}{\avg{1-\mu}}
    \simeq
    \frac{\int_0^2 \nu^2 h(\nu) \dd{\nu}}
         {\int_0^2 \nu h(\nu) \dd{\nu}}
    =
    \frac{1}{3}.
\end{equation}

\noindent Hence

\begin{equation}
    \avg{(1-\mu)^2}
    \simeq
    \frac{1-\avg{\mu}}{3},
\end{equation}

\noindent which yields

\begin{equation}
    \sigma_\mu
    \simeq
    \sqrt{\frac{1-\avg{\mu}}{3}} .
\end{equation}

\section{Dispersion relation for CR distributions with non-negligible pitch-angle or momentum spreads} \label{app:growth-rates-general}

Let us reconsider Sec.~\ref{sec:general-dispersion-relation} for a general gyrotropic CR distribution $f_{cr}(p,\mu)$. Now, zero net current implies a return current of 

\begin{equation}
    v_e = \frac{n_{cr}}{n_e}\avg{v_\parallel},
\end{equation}

\noindent where $v_\parallel = v\mu$ and 

\begin{equation}
    \avg{X} = \frac{2\pi}{n_{cr}} \int_0^\infty p^2 \dd{p} \int_{-1}^1 \dd{\mu} f_{cr} \, X
\end{equation}

\noindent denotes the zeroth angular moment (or expectation value) of quantity $X$. As long as $p^2 v f_{cr} \rightarrow 0$ as $p\rightarrow\infty$, we may integrate Eq.~\eqref{eq:cr-susceptibility-general} by parts and drop the boundary terms, yielding

\begin{multline}
    \chi_{cr}^{\sigma}
    =-\frac{\omega_{p,cr}^2}{\omega^2}\Bigg[
    \avg{\frac{1}{\Gamma}
      \frac{\omega-kv_\parallel}
      {\omega-kv_\parallel +\sigma\Omega_{cr}/\Gamma}}\\
      +\frac{1}{2}\avg{\frac{1}{\Gamma}
      \frac{(v_\perp^2/c^2)(k^2 c^2-\omega^2)}
      {(\omega-kv_\parallel+\sigma\Omega_{cr}/\Gamma)^2}}
    \Bigg],
\end{multline}

\noindent where $v_\perp = v \sqrt{1-\mu^2}$. This simplifies to the cold-beam susceptibility, Eq.~\eqref{eq:susceptibility}, when 

\begin{equation}
    f_{cr} = \frac{n_{cr}}{2\pi p_{cr}^2} \delta(p-p_{cr})\delta(\mu-1).
\end{equation}

Following identical steps to Sec.~\ref{sec:general-dispersion-relation}, the analogue of Eq.~\eqref{eq:dispersion-relation-factored} is then

\begin{multline} \label{eq:dispersion-relation-factored-general}
 k^2v_{A0}^2-\omega^2
 = \sigma\Omega_i\frac{n_{cr}}{n_i}
 \avg{\frac{(\omega-kv_\parallel)^2}
 {\omega-kv_\parallel+\sigma\Omega_{cr}/\Gamma}}\\
 -\frac{1}{2}\Omega_i^2\frac{n_{cr}}{n_i}
 \avg{\frac{1}{\Gamma}\frac{(v_\perp^2/c^2)(k^2c^2-\omega^2)}
 {(\omega-kv_\parallel
 +\sigma\Omega_{cr}/\Gamma)^2}}.
\end{multline}

\subsection{Non-resonant growth rate}

We take $\sigma = -1$ and assume the CRs are unmagnetized,

\begin{equation}
    \abs{\omega''} = \Gamma \abs{\omega - kv_\parallel} \gg \Omega_{cr},
\end{equation}

\noindent so that to 1\textsuperscript{st} order in $\Omega_{cr}/\omega''$, 

\begin{multline}\label{eq:non-res-dr-warm}
    k^2v_{A0}^2-\omega^2\simeq
 -\Omega_i\frac{n_{cr}}{n_i}
   \left(\omega-k\avg{v_\parallel}\right)
 -\Omega_i^2\frac{n_{cr}}{n_i}
   \avg{\frac{1}{\Gamma}}\\
 -\frac{1}{2}\Omega_i^2\frac{n_{cr}}{n_i}
 \avg{\frac{1}{\Gamma}\frac{(v_\perp^2/c^2)(k^2 c^2-\omega^2)}
 {(\omega-kv_\parallel)^2}}.
\end{multline} 

\noindent This is the exact same form as Eq.~\eqref{eq:non-res-dispersion-relation} except for an additional term quantifying the effect of CR gyration about the background field. Given $\abs{\omega} \ll k\abs{v_\parallel} \ll kc$, the last term becomes negligible relative to the penultimate term provided

\begin{equation}
    \avg{\frac{v_\perp^2}{\Gamma v_\parallel^2}} \ll 2 \avg{\frac{1}{\Gamma}}.
\end{equation}

A sufficient condition is simply for $v_\perp \ll \abs{v_\parallel}$ across the CR population. 

Assuming this, the results of Sec.~\ref{sec:nonres-dr} are identical under the substitutions

\begin{equation}
    v_{cr} \mapsto \avg{v_\parallel} \qc \frac{1}{\Gamma_{cr}} \mapsto \avg{\frac{1}{\Gamma}},
\end{equation}

\noindent which explains why properties of the non-resonant mode, in the unmagnetized limit, are predominantly determined by the return current rather than details of the CR distribution. 

\section{Relating the escaping CR current and shock distribution} \label{app:FEB-shock-relationship}

 Our analysis relies on relating the far-upstream escaping CR current to CR properties at the shock. Given the importance of this relation, we expand on previous treatments, aiming to make the underlying assumptions explicit \cite{Caprioli+2009,Bell+2013}. Unprimed quantities will denote CR properties as seen in the lab (shock) frame with coordinates $(\vb{x},\vb{p},t)$; $\vb{p}'$ will denote momentum in the ion bulk-rest frame, with $(\vb{x},\vb{p}',t)$ denoting coordinates in this \textit{mixed} frame. Furthermore, we denote the lab frame ion bulk velocity by $\vb{u}(\vb{x},t)$. Primes are dropped outside of this section and generally throughout the literature. 

In the lab frame, the Vlasov--Fokker--Planck (VFP) equation for the evolution of the CR distribution $f(\vb{x},\vb{p},t)$ has the conservative form

\begin{equation} \label{eq:VFP-lab}
    \pdv{f}{t} + \nabla_x\vdot\pqty{\vb{v}f} + \nabla_p\vdot\qty{e\pqty{\vb{E} + \frac{\vb{v}\cross\vb{B}}{c}}f} = C\bqty{f}.
\end{equation}

\noindent Here, $f$ denotes the non-thermal CR population with momentum above the DSA injection momentum, including confined and escaping CRs; $\vb{E}(\vb{x},t)$ and $\vb{B}(\vb{x},t)$ are the mean electromagnetic fields; and $C\bqty{\cdot}$ is a scattering operator arising from electromagnetic fluctuations $\delta \vb{E}$ and $\delta \vb{B}$ about the mean fields (e.g., pitch-angle scattering). If we assume the scattering operator redistributes CR momenta without changing the total CR particle number, then $\int_{\mathbb{R}^3} C\bqty{f} \dd[3]{p} = 0$. Subsequent integration over momentum space results in the lab frame conservation law

\begin{equation} \label{eq:conservation-law-lab}
    \pdv{n}{t} + \nabla_x\vdot\vb{F} = 0,
\end{equation}

\noindent where $n = \int_{\mathbb{R}^3} f \dd[3]{p}$ and $\vb{F} = n\avg{\vb{v}} = \int_{\mathbb{R}^3} \vb{v} f \dd[3]{p}$. 

Integrating Eq.~\eqref{eq:conservation-law-lab} over the shock upstream, $x\in\bqty{0^+,x_{\rm FEB}^-}$, in 1D planar geometry gives

\begin{equation}
    F_{\rm esc}(t) - F_{\rm sh}(t) = -\dv{t} \int_{0^+}^{x_{\rm FEB}^-} n \dd{x} \simeq 0,
\end{equation}

\noindent where $F_{\rm esc}(t) \equiv F_x(x_{\rm FEB}^-,t)$, $F_{\rm sh}(t) \equiv F_x(0^+,t)$, and the right-hand side is neglected in steady state. The same argument applied to any upstream interval shows that $F_x(x,t)$ is spatially uniform in steady state; in particular, $F_x(x_{\rm FEB}^+,t) = F_x(x_{\rm FEB}^-,t) = F_{\rm esc}(t)$, with the location of the FEB immaterial. What changes across this free-escape boundary is the flux's decomposition: escaping forward-beamed CRs have a markedly larger bulk velocity than the quasi-isotropic confined population, so the same flux is carried by a smaller number density beyond the boundary. The sharp FEB, with sides labeled $x_{\rm FEB}^-$ and $x_{\rm FEB}^+$, serves as a convenient idealization of what is, in reality, a smooth diffusive-to-ballistic transition.

To obtain the CR flux at the shock, we assume that $f'$ is gyrotropic about $x$ such that $f'(x,\vb{p}',t) = f'(x,p',\mu',t)$. Then, its Legendre expansion is given by

\begin{align} \label{eq:Legendre-expansion}
    f'(x,\vb{p}',t) &= \sum_{\ell=0}^\infty f'_\ell (x,p',t) P_\ell(\mu') \nonumber \\
    &= f'_0(x,p',t) + f'_1(x,p',t) \mu' + \cdots,
\end{align}

\noindent where $\mu' = p'_x/p'$, $f'_0$ is the monopole term, $f'_1$ the dipole, and so on. In the diffusive approximation---applicable near the shock front but not near the free-escape boundary---the $\ell$\textsuperscript{th} term in the Legendre expansion is smaller by a factor of $u/c$ than the $(\ell-1)$\textsuperscript{th}, which justifies dropping higher-order terms beyond the dipole \cite{Blandford+1987, Bell+2013}.

We note that the angular-integrated VFP equation under the diffusive approximation is given by Eq.~(11a) in Ref.~\cite{Bell+2013}; restated here under the steady-state assumption,

\begin{equation}\label{eq:steady-state-vfp}
    \pdv{x}\pqty{u_x f'_0 + \frac{v'}{3}f'_1} \simeq \pdv{u_x}{x} \frac{1}{3p'^2}\pdv{p'}\pqty{p'^3 f'_0}.
\end{equation}

From Eq.~\eqref{eq:Legendre-expansion}, it is straightforward to obtain

\begin{equation}
    n' = 4\pi \int_0^\infty f'_0 p'^2 \dd{p'} \qc F'_x = \frac{4\pi}{3}\int_0^\infty v' f'_1 p'^2 \dd{p'}
\end{equation}

\noindent by orthogonality of the Legendre polynomials. To relate lab and mixed frame flux, we note that $\dd[3]{p}/p^0$ is a Lorentz scalar and $f(\vb{x},\vb{p},t) = f'(\vb{x},\vb{p}',t)$, where $p^\mu = (\Gamma m c, \Gamma m \vb{v})$ is the CR four-momentum \cite{deGroot+80}. Hence, the Jacobian relating lab and mixed frames is

\begin{equation}
    \frac{\dd[3]{p}}{\dd[3]{p'}} = \frac{\Gamma}{\Gamma'}.
\end{equation}

Since the Lorentz boost of $p^\mu$ yields $\vb{p} \simeq \vb{p}' + \Gamma' m \vb{u}$ through $\order{u/c}$, we then have

\begin{equation} \label{eq:F-lab}
    \vb{F} = \int_{\mathbb{R}^3} \vb{v} f' \frac{\Gamma}{\Gamma'} \dd[3]{p'} \simeq \vb{F}' + \vb{u} n',
\end{equation}

\noindent and the lab flux becomes

\begin{equation} \label{eq:lab-flux}
    F_x = 4\pi \int_0^\infty \pqty{u_x f'_0 + \frac{v'}{3}f'_1} p'^2 \dd{p'}.
\end{equation}

Splitting this flux into bands above and below $p'_{\max}$, $F_x \equiv F_x^< + F_x^>$, and applying $4\pi \int_{p'_{\max}}^\infty \dd{p'} p'^2 (\cdot)$ to Eq.~\eqref{eq:steady-state-vfp} results in

\begin{equation}
    \pdv{F_x^>}{x} = -\frac{4\pi}{3} \pqty{p'_{\max}}^3 f'_0(x,p'_{\max}) \pdv{u_x}{x},
\end{equation}

\noindent where we assumed $p'^3 f_0' \rightarrow 0$ as $p' \rightarrow \infty$. Subsequent integration across the shock at $x\in\bqty{-\epsilon, \epsilon}$ as $\epsilon \rightarrow 0$ then yields

\begin{equation} \label{eq:pmax-flux-shock}
    F_x^> (0^+) - F_x^> (0^-) = \frac{4\pi}{q} v_{\rm sh} \pqty{p'_{\max}}^3 f'_0(0,p'_{\max}),
\end{equation}

\noindent where we used $u_x(0^+) = -v_{\rm sh}$, $u_x(0^-) = -v_{\rm sh}/r$, and $q = 3r/(r-1)$. The right-hand side represents the increase in flux of CRs with momentum $p'>p'_{\max}$ as they cross from downstream to upstream, which we denote by $S \geq 0$. The total lab frame upstream flux from the shock is then

\begin{align} \label{eq:Fsh}
    F_{\rm sh} &= F_x^<(0^+) + F_x^>(0^+) \nonumber\\
    &= F_x^{<}(0^+) + F_x^> (0^-) + \bqty{F_x^> (0^+) - F_x^> (0^-)} \nonumber\\
    &\simeq F_x^{<}(0^+) + F_x^> (0^-) + S \nonumber\\
    &\simeq S.
\end{align}

\noindent In the last line, $F_x^>(0^-) \simeq 0$ because there is essentially no downstream population with momentum $p' > p'_{\max}$, and Eq.~\eqref{eq:lab-flux} implies that $F_x^<(0^+) \simeq 0$ as long as the confined population's bulk-advection downstream cancels its anisotropic flux upstream. 

Combining our results, we have $F_{\rm esc} \simeq F_{\rm sh} \simeq S$. Since the mixed frame escaping CR current is

\begin{equation}
    J'_{\rm esc} = e F'_{\rm esc} \simeq eF_{\rm esc} \frac{\avg{v'_x}}{\avg{v'_x}-v_{\rm sh}},
\end{equation}

\noindent using Eq.~\eqref{eq:F-lab}, we obtain

\begin{align} 
    J'_{\rm esc} &= e n'_{\rm esc} \avg{v'_x}_{\rm esc} \nonumber \\
    &\simeq \frac{4\pi}{q} e v_{\rm sh} \pqty{p'_{\max}}^3 f'_0(0,p'_{\max}), \label{eq:Jcr-vshock-relation-general}
\end{align}

\noindent where $n'_{\rm esc}(t) \equiv n'(x_{\rm FEB}^+,t)$ and $\avg{v'_x}_{\rm esc}(t) \equiv \avg{v'_x}(x_{\rm FEB}^+,t)$, with the last equality neglecting terms of order $v_{\rm sh}/c$. Equation~\eqref{eq:Jcr-vshock-relation-general} gives the ion bulk-rest frame escaping CR current used in Eq.~\eqref{eq:Jcr-vshock-relation}.

\bibliography{references}

@article{Bell2004,
    author = {Bell, A. R.},
    title = {Turbulent amplification of magnetic field and diffusive shock acceleration of cosmic rays},
    journal = {Monthly Notices of the Royal Astronomical Society},
    volume = {353},
    number = {2},
    pages = {550-558},
    year = {2004},
    month = {09},
    issn = {0035-8711},
    doi = {10.1111/j.1365-2966.2004.08097.x},
}

@article{Holcomb+2019,
    doi = {10.3847/1538-4357/ab328a},
    year = {2019},
    month = {aug},
    publisher = {The American Astronomical Society},
    volume = {882},
    number = {1},
    pages = {3},
    author = {Holcomb, Cole and Spitkovsky, Anatoly},
    title = {On the Growth and Saturation of the Gyroresonant Streaming Instabilities},
    journal = {The Astrophysical Journal},
}

@article{Bell+2013,
    author = {Bell, A. R. and Schure, K. M. and Reville, B. and Giacinti, G.},
    title = {Cosmic-ray acceleration and escape from supernova remnants},
    journal = {Monthly Notices of the Royal Astronomical Society},
    volume = {431},
    number = {1},
    pages = {415-429},
    year = {2013},
    month = {02},
    issn = {0035-8711},
    doi = {10.1093/mnras/stt179},
}

@book{Gary1993, 
    place={Cambridge}, 
    series={Cambridge Atmospheric and Space Science Series}, 
    title={Theory of Space Plasma Microinstabilities}, 
    publisher={Cambridge University Press}, 
    author={Gary, S. Peter}, 
    year={1993}, 
    collection={Cambridge Atmospheric and Space Science Series},
}

@book{Melrose1986, 
    place={Cambridge}, 
    title={Instabilities in Space and Laboratory Plasmas}, 
    publisher={Cambridge University Press}, 
    author={Melrose, D. B.}, 
    year={1986},
}

@article{WinskeGary1986,
    author = {Winske, D. and Gary, S. P.},
    title = {Electromagnetic instabilities driven by cool heavy ion beams},
    journal = {Journal of Geophysical Research: Space Physics},
    volume = {91},
    number = {A6},
    pages = {6825-6832},
    doi = {10.1029/JA091iA06p06825},
    year = {1986},
}

@article{HenyeyGreenstein1941,
    author  = {Henyey, L. G. and Greenstein, J. L.},
    title   = {Diffuse Radiation in the {Galaxy}},
    journal = {The Astrophysical Journal},
    volume = {93},
    pages = {70--83},
    year = {1941},
    doi = {10.1086/144246},
}

@article{AmatoBlasi2009,
    author = {Amato, E. and Blasi, P.},
    title = {A kinetic approach to cosmic-ray-induced streaming instability at supernova shocks},
    journal = {Monthly Notices of the Royal Astronomical Society},
    volume = {392},
    number = {4},
    pages = {1591-1600},
    year = {2009},
    month = {02},
    doi = {10.1111/j.1365-2966.2008.14200.x},
}

@article{Haggerty+2019,
    doi = {10.3847/1538-4357/ab58c8},
    year = {2019},
    month = {dec},
    publisher = {The American Astronomical Society},
    volume = {887},
    number = {2},
    pages = {165},
    author = {Haggerty, Colby C. and Caprioli, Damiano},
    title = {{dHybridR}: A Hybrid Particle-in-cell Code Including Relativistic Ion Dynamics},
    journal = {The Astrophysical Journal},
}

@article{KulsrudPearce69,
    title={The effect of wave-particle interactions on the propagation of cosmic rays},
    author={Kulsrud, Russell and Pearce, William P},
    journal={The Astrophysical Journal},
    volume={156},
    pages={445},
    year={1969},
}

@book{Stix1992,
    title={Waves in Plasmas},
    author={Stix, Thomas H},
    year={1992},
    publisher={American Institute of Physics},
    address={New York},
}

@article{Lyutikov1999, 
    title={Beam instabilities in a magnetized pair plasma}, 
    volume={62}, 
    DOI={10.1017/S0022377899007837}, 
    number={1}, 
    journal={Journal of Plasma Physics}, 
    author={Lyutikov, Maxim}, 
    year={1999}, 
    pages={65--86},
}

@article{ONiel+68,
    author = {O'Neil, T. M. and Malmberg, J. H.},
    title = {Transition of the Dispersion Roots from Beam-Type to {Landau}-Type Solutions},
    journal = {The Physics of Fluids},
    volume = {11},
    number = {8},
    pages = {1754-1760},
    year = {1968},
    month = {08},
    issn = {0031-9171},
    doi = {10.1063/1.1692190},
}

@book{deGroot+80,
    author = {de Groot, S.~R. and van Leeuwen, W.~A. and van Weert, Ch.~G.},
    title = {Relativistic Kinetic Theory. Principles and Applications},
    year={1980},
    publisher={North-Holland Publishing Company},
}

@article{Lemmerz+2025,
    doi = {10.3847/1538-4357/ad8eb3},
    year = {2025},
    month = {jan},
    publisher = {The American Astronomical Society},
    volume = {979},
    number = {1},
    pages = {34},
    author = {Lemmerz, Rouven and Shalaby, Mohamad and Pfrommer, Christoph and Thomas, Timon},
    title = {The Theory of Resonant Cosmic Ray--driven Instabilities---Growth and Saturation of Single Modes},
    journal = {The Astrophysical Journal},
}

@article{Ellison+2011,
    doi = {10.1088/0004-637X/731/2/87},
    year = {2011},
    month = {mar},
    publisher = {The American Astronomical Society},
    volume = {731},
    number = {2},
    pages = {87},
    author = {Ellison, Donald C. and Bykov, Andrei M.},
    title = {GAMMA-RAY EMISSION OF ACCELERATED PARTICLES ESCAPING A SUPERNOVA REMNANT IN A MOLECULAR CLOUD},
    journal = {The Astrophysical Journal},
}

@article{Achterberg1983,
    title={Modification of scattering waves and its importance for shock acceleration},
    author={Achterberg, A},
    journal={Astronomy and Astrophysics},
    volume={119},
    pages={274--278},
    year={1983},
}

@article{Caprioli+2009,
    author = {Caprioli, D. and Blasi, P. and Amato, E.},
    title = {On the escape of particles from cosmic ray modified shocks},
    journal = {Monthly Notices of the Royal Astronomical Society},
    volume = {396},
    number = {4},
    pages = {2065-2073},
    year = {2009},
    month = {07},
    issn = {0035-8711},
    doi = {10.1111/j.1365-2966.2008.14298.x},
}

@article{Zirakashvili+2008,
    author = {Zirakashvili, Vladimir N. and Ptuskin, Vladimir S.},
    title = {The influence of the {Alfv\'enic} drift on the shape of cosmic ray spectra in {SNRs}},
    journal = {AIP Conference Proceedings},
    volume = {1085},
    number = {1},
    pages = {336-339},
    year = {2008},
    month = {12},
    issn = {0094-243X},
    doi = {10.1063/1.3076675},
}

@article{Ohira+2010,
	author = {Ohira, Y. and Murase, K. and Yamazaki, R.},
	title = {Escape-limited model of cosmic-ray acceleration revisited},
	DOI= "10.1051/0004-6361/200913495",
	journal = {Astronomy and Astrophysics},
	year = 2010,
	volume = 513,
	pages = "A17",
	month = "",
}

@article{Shalaby+2021,
    doi = {10.3847/1538-4357/abd02d},
    year = {2021},
    month = {feb},
    publisher = {The American Astronomical Society},
    volume = {908},
    number = {2},
    pages = {206},
    author = {Shalaby, Mohamad and Thomas, Timon and Pfrommer, Christoph},
    title = {A New Cosmic-Ray-driven Instability},
    journal = {The Astrophysical Journal},
}

@article{Weidl+2019,
    doi = {10.3847/1538-4357/ab0462},
    year = {2019},
    month = {mar},
    publisher = {The American Astronomical Society},
    volume = {873},
    number = {1},
    pages = {57},
    author = {Weidl, Martin S. and Winske, Dan and Niemann, Christoph},
    title = {Three Regimes and Four Modes for the Resonant Saturation of Parallel Ion-beam Instabilities},
    journal = {The Astrophysical Journal},
}

@article{Zacharegkas+2024,
    doi = {10.3847/1538-4357/ad3960},
    year = {2024},
    month = {may},
    publisher = {The American Astronomical Society},
    volume = {967},
    number = {1},
    pages = {71},
    author = {Zacharegkas, Georgios and Caprioli, Damiano and Haggerty, Colby and Gupta, Siddhartha and Schroer, Benedikt},
    title = {Modeling the Saturation of the {Bell} Instability Using Hybrid Simulations},
    journal = {The Astrophysical Journal},
}

@article{Caprioli+2010c,
    author = {Caprioli, D. and Kang, Hyesung and Vladimirov, A. E. and Jones, T. W.},
    title = {Comparison of different methods for non-linear diffusive shock acceleration},
    journal = {Monthly Notices of the Royal Astronomical Society},
    volume = {407},
    number = {3},
    pages = {1773--1783},
    year = {2010},
    month = {09},
    issn = {0035-8711},
    doi = {10.1111/j.1365-2966.2010.17013.x},
}

@ARTICLE{Malkov+2011,
    author = {{Malkov}, M.~A. and {Diamond}, P.~H. and {Sagdeev}, R.~Z.},
    title = "{Mechanism for spectral break in cosmic ray proton spectrum of supernova remnant W44}",
    journal = {Nature Communications},
    year = 2011,
    month = feb,
    volume = {2},
    eid = {194},
    pages = {194},
    doi = {10.1038/ncomms1195},
}

@article{Das+2026,
    doi = {10.3847/1538-4357/ae75ff},
    year = {2026},
    month = {jul},
    publisher = {The American Astronomical Society},
    volume = {1005},
    number = {2},
    pages = {155},
    author = {Das, Saikat and Gupta, Siddhartha and Sharma, Prateek},
    title = {Impact of Cosmic-Ray Distribution on the Growth and Saturation of {Bell} Instability},
    journal = {The Astrophysical Journal},
}

@article{Malkov+2013,
    doi = {10.1088/0004-637X/768/1/73},
    year = {2013},
    month = {apr},
    publisher = {The American Astronomical Society},
    volume = {768},
    number = {1},
    pages = {73},
    author = {Malkov, M. A. and Diamond, P. H. and Sagdeev, R. Z. and Aharonian, F. A. and Moskalenko, I. V.},
    title = {ANALYTIC SOLUTION FOR SELF-REGULATED COLLECTIVE ESCAPE OF COSMIC RAYS FROM THEIR ACCELERATION SITES},
    journal = {The Astrophysical Journal},
}

@article{Evoli+2018,
    title = {Self-generated cosmic-ray confinement in {TeV} halos: Implications for {TeV} $\ensuremath{\gamma}$-ray emission and the positron excess},
    author = {Evoli, Carmelo and Linden, Tim and Morlino, Giovanni},
    journal = {Physical Review D},
    volume = {98},
    issue = {6},
    pages = {063017},
    numpages = {11},
    year = {2018},
    month = {Sep},
    publisher = {American Physical Society},
    doi = {10.1103/PhysRevD.98.063017},
}

@article{Nava+2016,
    author = {Nava, L. and Gabici, S. and Marcowith, A. and Morlino, G. and Ptuskin, V. S.},
    title = {Non-linear diffusion of cosmic rays escaping from supernova remnants -- {I}. {The} effect of neutrals},
    journal = {Monthly Notices of the Royal Astronomical Society},
    volume = {461},
    number = {4},
    pages = {3552-3562},
    year = {2016},
    month = {10},
    issn = {0035-8711},
    doi = {10.1093/mnras/stw1592},
}

@article{Nava+2019,
    author = {Nava, L and Recchia, S and Gabici, S and Marcowith, A and Brahimi, L and Ptuskin, V},
    title = {Non-linear diffusion of cosmic rays escaping from supernova remnants -- {II}. {Hot} ionized media},
    journal = {Monthly Notices of the Royal Astronomical Society},
    volume = {484},
    number = {2},
    pages = {2684-2691},
    year = {2019},
    month = {04},
    issn = {0035-8711},
    doi = {10.1093/mnras/stz137},
}

@article{Schroer+2021,
    doi = {10.3847/2041-8213/ac02cd},
    year = {2021},
    month = {jun},
    publisher = {The American Astronomical Society},
    volume = {914},
    number = {1},
    pages = {L13},
    author = {Schroer, Benedikt and Pezzi, Oreste and Caprioli, Damiano and Haggerty, Colby and Blasi, Pasquale},
    title = {Dynamical Effects of Cosmic Rays on the Medium Surrounding Their Sources},
    journal = {The Astrophysical Journal Letters},
}

@article{Cermenati+2026,
	author = {Cermenati, Alessandro and Aloisio, Roberto and Blasi, Pasquale and Evoli, Carmelo},
	title = {Excitation of the nonresonant streaming instability around sources of ultrahigh-energy cosmic rays},
	DOI= "10.1051/0004-6361/202556040",
	journal = {Astronomy and Astrophysics},
	year = 2026,
	volume = 707,
	pages = "A19",
}

@article{Blandford+1987,
    title = {Particle acceleration at astrophysical shocks: A theory of cosmic ray origin},
    journal = {Physics Reports},
    volume = {154},
    number = {1},
    pages = {1-75},
    year = {1987},
    issn = {0370-1573},
    doi = {https://doi.org/10.1016/0370-1573(87)90134-7},
    author = {Roger Blandford and David Eichler},
}

@article{Bell1978,
    author = {Bell, A. R.},
    title = {The acceleration of cosmic rays in shock fronts -- {I}},
    journal = {Monthly Notices of the Royal Astronomical Society},
    volume = {182},
    number = {2},
    pages = {147-156},
    year = {1978},
    month = {02},
    issn = {0035-8711},
    doi = {10.1093/mnras/182.2.147},
}

@article{Blandford+1978,
	Author = {Blandford, R.~D. and Ostriker, J.~P.},
	Doi = {10.1086/182658},
	Journal = {The Astrophysical Journal Letters},
	Month = apr,
	Pages = {L29-L32},
	Title = {Particle acceleration by astrophysical shocks},
	Volume = {221},
	Year = {1978},
}

@article{Blasi2019,
	Author = {Blasi, Pasquale},
	Journal = {La Rivista del Nuovo Cimento},
	Month = dec,
	Number = {12},
	Pages = {549-600},
	Title = {Acceleration of galactic cosmic rays},
	Volume = {42},
	Year = 2019,
}

@article{Caprioli2012,
    doi = {10.1088/1475-7516/2012/07/038},
    year = {2012},
    month = {jul},
    publisher = {},
    volume = {2012},
    number = {07},
    pages = {038},
    author = {Damiano Caprioli},
    title = {Cosmic-ray acceleration in supernova remnants: non-linear theory revised},
    journal = {Journal of Cosmology and Astroparticle Physics},
}

@article{Caprioli+2020,
    doi = {10.3847/1538-4357/abbe05},
    year = {2020},
    month = {dec},
    publisher = {The American Astronomical Society},
    volume = {905},
    number = {1},
    pages = {2},
    author = {Caprioli, Damiano and Haggerty, Colby C. and Blasi, Pasquale},
    title = {Kinetic Simulations of Cosmic-Ray-modified Shocks. {II}. {Particle} Spectra},
    journal = {The Astrophysical Journal},
}

@inproceedings{Haggerty+2019p,
    Archiveprefix = {arXiv},
    Author = {Haggerty, C. and Caprioli, D. and Zweibel, E.},
    Booktitle = {36th International Cosmic Ray Conference (ICRC2019)},
    Eid = {279},
    Eprint = {1909.06346},
    Month = jul,
    Pages = {279},
    Primaryclass = {astro-ph.HE},
    Series = {International Cosmic Ray Conference},
    Title = {{Hybrid Simulations of the Resonant and Non-Resonant Cosmic Ray Streaming Instability}},
    Volume = {36},
    Year = {2019}
}

@article{Blasi+2015,
    title = {High-Energy Cosmic Ray Self-Confinement Close to Extra-Galactic Sources},
    author = {Blasi, Pasquale and Amato, Elena and D'Angelo, Marta},
    journal = {Phys. Rev. Lett.},
    volume = {115},
    issue = {12},
    pages = {121101},
    numpages = {5},
    year = {2015},
    month = {Sep},
    publisher = {American Physical Society},
    doi = {10.1103/PhysRevLett.115.121101},
}

@inproceedings{Zweibel1979,
	author = {Zweibel, Ellen G.},
	Booktitle = {Particle Acceleration Mechanisms in Astrophysics},
	Doi = {10.1063/1.32090},
	Editor = {Arons, J. and McKee, C. and Max, C.},
	Month = nov,
	Pages = {319-328},
	Series = {American Institute of Physics Conference Series},
	Title = {{Energetic particle trapping by Alfven wave instabilities}},
	Volume = {56},
	Year = 1979,
}

\end{document}